\documentclass[11pt]{article}
\usepackage[utf8]{inputenc}
\usepackage[margin=1in]{geometry}
\usepackage{amsmath,amssymb,amsthm,mathtools}
\usepackage[shortlabels]{enumitem}
\usepackage{microtype,graphicx,booktabs,xcolor}
\usepackage[font=small,labelfont=bf]{caption}
\usepackage[colorlinks=true,linkcolor=blue!50!black,citecolor=green!40!black,urlcolor=blue!50!black]{hyperref}
\graphicspath{{figures/}}

\renewcommand{\thesection}{\Roman{section}}

\makeatletter
\renewcommand{\p@subsection}{\thesection.}
\makeatother

\makeatletter\@addtoreset{equation}{section}\makeatother
\usepackage{etoolbox}
\makeatletter
\patchcmd{\l@section}{1.5em}{2.6em}{}{}
\renewcommand*\l@subsection{\@dottedtocline{2}{2.6em}{2.0em}}
\makeatother

\newtheorem{theorem}{Theorem}
\newtheorem{proposition}[theorem]{Proposition}
\newtheorem{lemma}[theorem]{Lemma}
\newtheorem{corollary}[theorem]{Corollary}

\newcommand{\R}{\mathbb{R}}
\newcommand{\C}{\mathbb{C}}
\newcommand{\dmuB}{{\rm d}\mu_B}
\newcommand{\dmuL}{{\rm d}\mu_L}
\newcommand{\s}{\mathfrak{s}}
\newcommand{\vk}{{\vec k}}
\newcommand{\Fn}[1]{\lVert #1\rVert_{\mathcal F}}

\title{Classical phase space distributions for quantum states in cosmology:\\[4pt]
\large general states, operator ordering and the geometry of fading non-commutativity}
\author{Alejandro Corichi\\[2pt]
\small Centro de Ciencias Matem\'aticas, Universidad Nacional Aut\'onoma de M\'exico,\\
\small UNAM-Campus Morelia, Morelia, Michoac\'an, Mexico\\
\small{\tt corichi@matmor.unam.mx}}
\date{}

\begin{document}
\maketitle

\begin{abstract}
In earlier work with Ashtekar and Kesavan we showed that, for linear fields in any FLRW background, the vacuum defines a probability distribution $\rho_o$ on the classical phase space whose moments are exactly the Weyl-ordered $n$-point functions. We extend that analysis in four directions. First, for an arbitrary state $\Psi$ we give an explicit formula, in the Bargmann representation, for the distribution $\rho_\Psi$ that reproduces all Weyl-ordered expectation values. For pure states it is non-negative precisely for coherent and squeezed states, each the Gaussian of a compatible K\"ahler metric, and the Gaussian smoothing needed to make it positive is either zero or the full coherent-state width. Second, for linear observables every operator ordering differs from the Weyl ordering by explicit products of $c$-number commutators. Third, for two linear observables in a pure Gaussian state, the classical correlation coefficient, the ratio of commutator to anticommutator and the squeezing are functions of a single angle between their Hamiltonian vector fields; this bounds the departure of arbitrarily ordered $n$-point functions from the moments of $\rho$. Fourth, for fields in FLRW the equality holds at arbitrary, possibly different, times and for every homogeneous isotropic quasi-free state. In de Sitter the non-commutative corrections fall as $e^{-N}$ for field and momentum and as $e^{-3N}/3$ for the field at two times. For the state inherited from inflation they stay suppressed in the radiation era, by the inverse occupation number, except near the nodes of the mode functions. The same squeezing entangles the modes $\pm\vec k$ relative to the radiation-era vacuum and makes displaced-parity measurements violate Bell inequalities, so the positive distribution describes a specified family of observables, not all measurements.
\end{abstract}

\tableofcontents
\section{Introduction}\label{sec:intro}

In the cosmological literature, the emergence of classical behavior of primordial perturbations is usually attributed to three properties of the quantum state of linear perturbations during inflation: the fading of quantum non-commutativity, the large squeezing of the state, and the possibility of replacing the quantum state by a distribution function on the classical phase space \cite{GrishchukSidorov,AFJP,PolarskiStarobinsky,KPS,KieferPolarski}. The role of squeezing in this account has been questioned by Agullo, Bonga and Ribes Metidieri, who argue that in a generic FLRW background the squeezing of the perturbations is as ambiguous as particle creation, that once the ambiguity is resolved with the symmetries of de Sitter space inflation does not squeeze them, and that the perturbations remain quantum as long as non-Gaussianity can be neglected \cite{ABRM}. In \cite{ACK}, Ashtekar, Kesavan and the present author re-examined these three notions in general FLRW space-times, using the geometric structures on the covariant phase space $\Gamma$ of solutions: the symplectic structure $\Omega$, the complex structure $J$ that selects the vacuum, and the Riemannian metric $g(\phi_1,\phi_2)=\Omega(\phi_1,J\phi_2)$. We found that the three notions are conceptually distinct, that inflation is not essential for any of them, and that the third one is realized in a surprisingly strong sense. In the Bargmann representation the vacuum $\Psi_o$ is associated with a normalized distribution $\rho_o$ on $\Gamma$, and the vacuum expectation value of every Weyl-ordered polynomial in the field and momentum operators equals the classical expectation value, with respect to $\rho_o$, of the corresponding phase space function. The same is true for coherent and squeezed states.

That result raises several questions which were left open in \cite{ACK}, and which this paper addresses.
\begin{enumerate}[(a)]
\item \emph{Which states?} Is there a distribution for states other than coherent and squeezed states, how is it obtained from the Bargmann wave function, and when is it a genuine (non-negative) probability distribution? What does the Bargmann density $|\Psi|^2\dmuB$ itself represent when $\Psi$ is not the vacuum? What happens for superpositions?
\item \emph{Which orderings?} The quantum and classical predictions differ only for products of operators that are not Weyl ordered, and the reordering terms vanish as $\hbar\to0$ \cite{ACK}. Can these terms be given exactly, and how large are they compared with the classical moments, when $\hbar$ is fixed?
\item \emph{How are the three notions related?} The fading of non-commutativity was measured in \cite{ACK} by the ratio $R_{\varphi,\pi}$ of commutator to anticommutator, and squeezing by the norms of the Hamiltonian vector fields $X_{\varphi_\vk}$ and $X_{\pi_\vk}$. Both decrease or grow during inflation, while $\rho_o$ is the same at all times. Are these quantities related to $\rho_o$?
\item \emph{Fields.} Does the equality of $n$-point functions extend to operators at different times and to other homogeneous isotropic states, and what happens after inflation, when the modes re-enter the Hubble radius?
\end{enumerate}

Our main results are the following.

\begin{enumerate}[(i)]
\item \emph{The distribution of an arbitrary state} (Sec.~\ref{sec:general}). For any state $\Psi$ in the Bargmann representation we give an explicit formula, Eq.~\eqref{eq:rhoPsi}, for a real normalized function $\rho_\Psi$ on $\Gamma$ such that $\langle(\hat x^n\hat p^m)_W\rangle_\Psi=\langle x^np^m\rangle_{\rho_\Psi}$ for all $m,n$. It is the Wigner function of $\Psi$, written directly in the representation used in \cite{ACK}. Whether it is a probability distribution is decided by positive definiteness of the generating function $\langle\hat W(\lambda,\mu)\rangle_\Psi$, and a finite test with four Weyl operators already detects the failure for the first excited state.
\item \emph{Positivity} (Sec.~\ref{sec:positivity}). $\rho_\Psi\ge0$ for a pure state if and only if $\Psi(\zeta)=C\,e^{\s\zeta^2+\beta\zeta}$ with $|\s|<\frac12$. The coherent and squeezed states of \cite{ACK} therefore exhaust the pure states admitting a classical distribution. Each such $\rho_\Psi$ is the Gaussian $\pi^{-1}e^{-g_\s(\zeta-\zeta_c,\zeta-\zeta_c)}$ of a K\"ahler metric $g_\s$ compatible with $\Omega$, centred at a point $\zeta_c$ of $\Gamma$.
\item \emph{The Bargmann density} (Sec.~\ref{sec:bargmanndensity}). $|\Psi|^2\dmuB$ is positive for every state and reproduces exactly the anti-normally ordered products of $\hat a$ and $\hat a^\dagger$, Eq.~\eqref{eq:antinormal}. It coincides with $\rho_\Psi$ for the vacuum only; for a coherent state the two are centred at different points. Gaussian smoothing interpolates between the two (Sec.~\ref{sec:smoothing}). For a pure state the smoothing needed to make the distribution positive is either zero, for Gaussian states, or the full coherent-state width: there is nothing in between.
\item \emph{Superpositions} (Sec.~\ref{sec:superpositions}). For $\Psi=\mathcal N(c_1\Psi_{\zeta_1}+c_2\Psi_{\zeta_2})$ we give $\rho_\Psi$ in closed form. The difference from the corresponding mixture is a localized interference term, invisible to every observable whose resolution in phase space is coarser than the separation of the packets, Eq.~\eqref{eq:superbound}.
\item \emph{Arbitrary orderings} (Sec.~\ref{sec:ordering}). For linear observables $\hat L_i$, the product $\hat L_1\cdots\hat L_n$ equals the sum, over all sets of disjoint pairs, of products of half-commutators times Weyl-ordered products of the remaining factors, Eq.~\eqref{eq:ordering}. Hence $\rho_\Psi$ and the $c$-number commutators determine every expectation value.
\item \emph{One angle} (Sec.~\ref{sec:angle}). For two linear observables $F,G$ in a pure Gaussian state, let $\theta_{FG}$ be the angle between $X_F$ and $X_G$ in the K\"ahler metric of the state. Then the correlation coefficient of $F$ and $G$ in $\rho$ is $\cos\theta_{FG}$, the relative size of their commutator is $|\sin\theta_{FG}|$, the ratio $R$ of \cite{ACK} is $|\tan\theta_{FG}|$, and for a canonical pair $\|X_F\|\,\|X_G\|=1/|\sin\theta_{FG}|$. For general states an inequality holds. As a consequence, the departure of any arbitrarily ordered $n$-point function from the corresponding moment of $\rho$ is bounded by an explicit multiple of the largest $|\sin\theta|$, Eq.~\eqref{eq:npointbound}.
\item \emph{Fields in FLRW} (Sec.~\ref{sec:fields}). The distribution $\rho_o$ lives on the covariant phase space and does not evolve. The time dependence is carried by the classical observables $\varphi_\vk(\eta)$ and $\pi_\vk(\eta)$. As a result, Weyl-ordered $n$-point functions agree with the moments of $\rho_o$ at arbitrary, possibly different, times, and for mixed as well as pure homogeneous isotropic quasi-free states, extending the equal-time result for quasi-free vacua of \cite{ACK}. In de Sitter the relative size of the non-commutative corrections is $[1+(aH/k)^2]^{-1/2}\simeq e^{-N}$ for $\varphi$ and $\pi$ at equal times, and $\simeq e^{-3N}/3$ for $\varphi$ at two times, which gives the two-time ratio of \cite[App.~A]{ACK} in closed form. In a radiation era that follows inflation, with the state inherited from inflation, it is of order $1/(4n_\vk)$ after re-entry, where $n_\vk$ is the occupation number with respect to the radiation-era vacuum. The exceptions are windows of relative duration $\sim1/(4n_\vk)$ around the nodes of the mode functions.
\item \emph{What is not captured} (Sec.~\ref{sec:notcaptured}). In the same state, the modes $\pm\vk$ form a two-mode squeezed vacuum with respect to the radiation-era vacuum, with $\sinh^2r_k=n_\vk$. Their entanglement entropy is $\simeq4N_k$, so their quantum mutual information is $\simeq8N_k$. The classical mutual information of $\rho_o$ grows at the same rate but is a different quantity, the R\'enyi-2 mutual information of the state, smaller than the von Neumann one by $2-2\ln2$ at late times. Displaced-parity measurements violate the CHSH inequality, approaching $8\cdot3^{-9/8}\simeq2.32$. The squeezing that makes linear observables classical makes these observables maximally non-classical. The modes $\pm\vk$ are global; between field modes of compact support, which local observers can access, the Bunch--Davies state of de Sitter space carries less entanglement than the Minkowski vacuum \cite{RAB,RABlong}.
\end{enumerate}

The upshot can be summarized as follows. The existence of a classical distribution reproducing Weyl-ordered $n$-point functions is a property of all pure Gaussian states, and of no other pure states. It holds at all times for linear dynamics, whether or not there is inflation. Inflation is what makes the remaining, non-Weyl-ordered information negligible, and the measure of this negligibility, $|\sin\theta|$, is a property of the distribution $\rho_o$ itself, read through the time-dependent observables. In this sense the three notions of \cite{ACK} remain conceptually distinct but become quantitatively tied together.

The paper is organized as follows. Section~\ref{sec:general} treats a single degree of freedom in the notation of \cite[Sec.~V.A]{ACK}. Section~\ref{sec:ordering} discusses arbitrary orderings and the geometric angle. Section~\ref{sec:fields} turns to fields in FLRW space-times, with de Sitter and radiation-dominated examples, and discusses the correlations between the modes $\pm\vk$ that the positive distribution does not capture. Section~\ref{sec:discussion} contains a discussion. We follow the notation and conventions of \cite{ACK} throughout, and refer to its equations as, e.g., \cite[(5.15)]{ACK}.
\section{A single degree of freedom: general states}\label{sec:general}

\subsection{The setting of \texorpdfstring{\cite{ACK}}{[1]}}\label{sec:setting}

We use the conventions of \cite[Sec.~V.A]{ACK}. We set $\hbar=1$ in this section and the next; physical units are restored as in \cite[Remark (1), Sec.~V.A]{ACK}. The canonical operators satisfy $[\hat x,\hat p]=i$, and $\hat a=(\hat x+i\hat p)/\sqrt2$. The phase space $\Gamma=\R^2$ carries the complex coordinate $\zeta=x+ip$, the Liouville measure $\dmuL={\rm d}x\,{\rm d}p$, and the normalized Bargmann measure
\begin{equation}\label{eq:muB}
\dmuB=\frac1\pi\,e^{-\zeta\bar\zeta}\,\dmuL .
\end{equation}
States are holomorphic functions $\Psi(\zeta)$ with $\langle\Phi,\Psi\rangle=\int_\Gamma\overline{\Phi(\zeta)}\,\Psi(\zeta)\,\dmuB$, and
\begin{equation}\label{eq:aadag}
\hat a^\dagger\Psi(\zeta)=\zeta\,\Psi(\zeta),\qquad \hat a\Psi(\zeta)=\frac{{\rm d}\Psi}{{\rm d}\zeta}(\zeta).
\end{equation}
The vacuum is $\Psi_o(\zeta)=1$, the functions $\Psi_n(\zeta)=\zeta^n/\sqrt{n!}$ form an orthonormal basis, and the reproducing property
\begin{equation}\label{eq:reproducing}
\int_\Gamma\overline{e^{\bar c\,\zeta}}\;h(\zeta)\,\dmuB=h(c)
\end{equation}
holds for every $h$ in the Bargmann space and every $c\in\C$. The Weyl operators and their classical counterparts are
\begin{equation}\label{eq:weyl}
\hat W(\lambda,\mu)=e^{i(\lambda\hat x+\mu\hat p)}=e^{-\frac12\bar\alpha\alpha}\,e^{-\bar\alpha\hat a^\dagger}e^{\alpha\hat a},\qquad W(\lambda,\mu)=e^{i(\lambda x+\mu p)},\qquad \alpha=\frac{\mu+i\lambda}{\sqrt2},
\end{equation}
and $\hat W(\lambda,\mu)\Psi(\zeta)=e^{-\frac12\alpha\bar\alpha}e^{-\bar\alpha\zeta}\Psi(\zeta+\alpha)$. Derivatives of $\langle\hat W(\lambda,\mu)\rangle$ at the origin generate expectation values of Weyl-ordered products $(\hat x^n\hat p^m)_W$, the average over the $\binom{n+m}{n}$ distinct orderings, while derivatives of $\langle W(\lambda,\mu)\rangle_\rho$ generate the moments $\langle x^np^m\rangle_\rho$ of a distribution $\rho$. For the vacuum, $\langle\hat W(\lambda,\mu)\rangle_o=e^{-\frac14(\lambda^2+\mu^2)}$, which coincides with the classical expectation value in
\begin{equation}\label{eq:rhoo}
\rho_o=\frac1\pi\,e^{-\zeta\bar\zeta},\qquad \int_\Gamma\rho_o\,\dmuL=1 ,
\end{equation}
so that $\langle(\hat x^n\hat p^m)_W\rangle_o=\langle x^np^m\rangle_{\rho_o}$ for all $m,n$ \cite[(5.15)]{ACK}. The exponent in $\rho_o$ is the norm of $\zeta$ in the metric $g$ defined by $J$ and $\Omega$. In \cite{ACK} the same equality was established for coherent states, and for squeezed states $\Psi(\zeta)=e^{\s\zeta^2}$ through a linear canonical transformation. We now consider arbitrary states.

\subsection{The distribution of an arbitrary state}\label{sec:rhoPsi}

The argument of \cite{ACK} uses only the equality of the two generating functions. For a general normalized state $\Psi$ we therefore define $\rho_\Psi$ as the inverse Fourier transform of the quantum one,
\begin{equation}\label{eq:rhodef}
\rho_\Psi(x,p)=\frac1{(2\pi)^2}\int_{\R^2}e^{-i(\lambda x+\mu p)}\,\langle\hat W(\lambda,\mu)\rangle_\Psi\,{\rm d}\lambda\,{\rm d}\mu .
\end{equation}
This is the Wigner function of $\Psi$ \cite{Wigner,Moyal,Hillery}. The point of this subsection is that it has a simple expression in the Bargmann representation, which makes the construction of \cite{ACK} available for every state.

\begin{proposition}\label{prop:rhoPsi}
Let $\Psi$ be a normalized state in the Bargmann representation. Then:
\begin{enumerate}[(a)]
\item $\rho_\Psi$ is real, bounded by $|\rho_\Psi|\le1/\pi$, and $\int_\Gamma\rho_\Psi\,\dmuL=1$;
\item $\langle\hat W(\lambda,\mu)\rangle_\Psi=\langle W(\lambda,\mu)\rangle_{\rho_\Psi}$ for all $\lambda,\mu$; consequently
\begin{equation}\label{eq:equalityPsi}
\langle(\hat x^n\hat p^m)_W\rangle_\Psi=\langle x^np^m\rangle_{\rho_\Psi}\qquad\text{for all }m,n,
\end{equation}
whenever the moments exist;
\item $\rho_\Psi$ is given by
\begin{equation}\label{eq:rhoPsi}
\rho_\Psi(\zeta)=\frac1\pi\,e^{-\zeta\bar\zeta}\int_\Gamma\overline{\Psi(w)}\;e^{\sqrt2\,\zeta w}\;\Psi\bigl(\sqrt2\,\bar\zeta-w\bigr)\,\dmuB(w).
\end{equation}
\end{enumerate}
\end{proposition}

\begin{proof}
Let $\hat D(\beta)=e^{\beta\hat a^\dagger-\bar\beta\hat a}$ and let $\hat\Pi\Psi(\zeta)=\Psi(-\zeta)$ be the parity operator. It is a standard fact \cite{Royer,CahillGlauber} that the inverse Fourier transform \eqref{eq:rhodef} equals $\pi^{-1}\langle\Psi,\hat D(\beta)\hat\Pi\hat D(\beta)^\dagger\Psi\rangle$, where $\beta=(x+ip)/\sqrt2=\zeta/\sqrt2$ is the eigenvalue of $\hat a$ on the coherent state peaked at $\zeta$. Since $\hat D(\beta)\hat\Pi\hat D(\beta)^\dagger$ is self-adjoint and unitary, (a) follows, together with normalization, which is $\langle\hat W(0,0)\rangle_\Psi=1$. Part (b) is Fourier inversion. For (c), \eqref{eq:aadag} gives $\bigl(\hat D(\beta)\Psi\bigr)(w)=e^{-\frac12\beta\bar\beta}e^{\beta w}\Psi(w-\bar\beta)$, and therefore
\begin{equation*}
\bigl(\hat D(\beta)\hat\Pi\hat D(\beta)^\dagger\Psi\bigr)(w)=e^{-2\beta\bar\beta}\,e^{2\beta w}\,\Psi(2\bar\beta-w).
\end{equation*}
Taking the inner product with $\Psi$ and setting $\beta=\zeta/\sqrt2$ gives \eqref{eq:rhoPsi}.
\end{proof}

Formula \eqref{eq:rhoPsi} is easily evaluated with \eqref{eq:reproducing}. For the vacuum, the integral equals one and we recover \eqref{eq:rhoo}. For the first excited state $\Psi_1=\zeta$, the integral is the derivative at $w=0$ of $e^{\sqrt2\zeta w}(\sqrt2\bar\zeta-w)$, so
\begin{equation}\label{eq:rho1}
\rho_{\Psi_1}(\zeta)=\frac1\pi\,e^{-\zeta\bar\zeta}\,\bigl(2\zeta\bar\zeta-1\bigr).
\end{equation}
The equality \eqref{eq:equalityPsi} holds for $\Psi_1$, but $\rho_{\Psi_1}$ is negative in the disc $\zeta\bar\zeta<\frac12$, so it is not a probability distribution. For the normalized coherent state $\Psi_{\zeta_o}=e^{-\frac12\bar\zeta_o\zeta_o}e^{\zeta_o\zeta}$, two applications of \eqref{eq:reproducing} give
\begin{equation}\label{eq:rhocoh}
\rho_{\Psi_{\zeta_o}}(\zeta)=\frac1\pi\,e^{-(\zeta-\sqrt2\zeta_o)(\bar\zeta-\sqrt2\bar\zeta_o)} .
\end{equation}
This is the vacuum distribution translated to the point $\zeta=\sqrt2\,\zeta_o$, at which $\Psi_{\zeta_o}$ is peaked \cite[(5.4)]{ACK}.\footnote{With the normalization of \cite[(5.4)]{ACK}, in which $\langle\Psi_{\zeta_o},(\hat x+i\hat p)\Psi_{\zeta_o}\rangle=\sqrt2\,\zeta_o$, the distribution reproducing \cite[(5.17)]{ACK} is centred at $\sqrt2\,\zeta_o$; equivalently, $\zeta_o$ in \cite[(5.16)]{ACK} should be read as the centre of the packet in the variable $\zeta$.} Since $\rho_\Psi$ is linear in the projector $|\Psi\rangle\langle\Psi|$, the same construction applies to mixed states: a density operator $\sum_ip_i|\Psi_i\rangle\langle\Psi_i|$ has distribution $\sum_ip_i\rho_{\Psi_i}$, and \eqref{eq:equalityPsi} holds with $\langle\cdot\rangle_\Psi$ replaced by the trace.

Thus the construction of \cite{ACK} extends to every state: the Weyl-ordered expectation values are always moments of a real normalized function on the classical phase space. What is not automatic is positivity. The generating-function argument of \cite{ACK} already contains the criterion. By Bochner's theorem \cite{Bochner}, $\rho_\Psi$ is a probability distribution if and only if $\langle\hat W(\lambda,\mu)\rangle_\Psi$ is the characteristic function of one, i.e.\ if and only if the matrices
\begin{equation}\label{eq:bochner}
M_{ij}=\bigl\langle\hat W(\lambda_j-\lambda_i,\mu_j-\mu_i)\bigr\rangle_\Psi
\end{equation}
are positive semidefinite for every finite set of points $(\lambda_i,\mu_i)$. Every quantum state satisfies a different, twisted condition. Positivity of $\langle\hat A^\dagger\hat A\rangle_\Psi$ for $\hat A=\sum_jc_j\hat W(\lambda_j,\mu_j)$ gives positivity of $e^{\frac i2(\lambda_i\mu_j-\mu_i\lambda_j)}M_{ij}$, and the Weyl phase cannot in general be removed \cite{DangniamFerrie}. A single finite test therefore suffices to show that no probability distribution reproduces all Weyl-ordered expectation values. For $\Psi_1$, $\langle\hat W(\lambda,\mu)\rangle=e^{-\frac14(\lambda^2+\mu^2)}\bigl(1-\frac12(\lambda^2+\mu^2)\bigr)$. The four points $(\pm\sqrt3,0)$, $(0,\pm\sqrt3)$ give a matrix \eqref{eq:bochner} with eigenvalue $1-4e^{-3/2}-5e^{-3}\simeq-0.1415$, while the twisted matrix is positive. Equivalently, the non-negative function $[\cos(\sqrt3x)+\cos(\sqrt3p)]^2$ has a negative Weyl-ordered expectation value in $\Psi_1$. Covariances alone cannot detect this: $\Psi_1$ has covariance $\frac32$ times the identity, which is also the covariance of a Gaussian probability distribution, that of a thermal state. Before doing so, note that the example \eqref{eq:rho1} also clarifies the remark in \cite{ACK} that, in the configuration representation, one can reproduce expectation values of functions of $\hat x$ alone by $|\psi(x)|^2\sigma(p)$ with an arbitrary $\sigma$. Once both $\hat x$ and $\hat p$ are involved, the distribution is unique, because \eqref{eq:rhodef} is determined by the expectation values of all Weyl operators. When $\rho_\Psi$ is negative somewhere, no probability distribution reproduces all Weyl-ordered expectation values.

\subsection{Which pure states admit a classical distribution}\label{sec:positivity}

For pure states the answer is complete, and it shows that the cases considered in \cite{ACK} are not just examples.

\begin{theorem}\label{thm:hudson}
For a normalized state $\Psi$ in the Bargmann representation, $\rho_\Psi\ge0$ if and only if
\begin{equation}\label{eq:gaussianPsi}
\Psi(\zeta)=C\,e^{\s\zeta^2+\beta\zeta},\qquad \s,\beta,C\in\C,\quad |\s|<\tfrac12 .
\end{equation}
\end{theorem}

This is Hudson's theorem \cite{Hudson} (see \cite{SotoClaverie} for several degrees of freedom), stated in the Bargmann language of \cite{ACK}. The proof is particularly transparent in this representation, so we recall it. Suppose $\rho_\Psi\ge0$. The Bargmann density $\pi^{-1}e^{-\zeta\bar\zeta}|\Psi(\zeta)|^2$ is a Gaussian average of $\rho_\Psi$, Eq.~\eqref{eq:BfromRho} below; since $\rho_\Psi$ is non-negative and normalized and the Gaussian is strictly positive, the average is strictly positive, so $\Psi$ has no zeros. The reproducing kernel gives $|\Psi(\zeta)|\le\|\Psi\|\,e^{\zeta\bar\zeta/2}$, so $\Psi$ is an entire function of order at most two. By Hadamard's factorization theorem a zero-free entire function of order at most two is the exponential of a polynomial of degree at most two, which is \eqref{eq:gaussianPsi}. Normalizability requires $|\s|<\frac12$. Conversely, Proposition~\ref{prop:squeezed} below shows that every such state has a Gaussian, hence positive, distribution.

\begin{proposition}\label{prop:squeezed}
For $\Psi$ as in \eqref{eq:gaussianPsi},
\begin{equation}\label{eq:rhosqueezed}
\rho_\Psi(\zeta)=\frac1\pi\exp\bigl[-g_\s(\zeta-\zeta_c,\zeta-\zeta_c)\bigr],\qquad g_\s(\zeta,\zeta)=\frac{(1+4\s\bar\s)\,\zeta\bar\zeta-4\,{\rm Re}(\s\bar\zeta^2)}{1-4\s\bar\s},
\end{equation}
with centre
\begin{equation}\label{eq:centre}
\zeta_c=\frac{\sqrt2\,(\beta+2\s\bar\beta)}{1-4\s\bar\s}.
\end{equation}
The quadratic form $g_\s$ is positive definite with unit determinant in the coordinates $(x,p)$. Hence $g_\s(u,v)=\Omega(u,J_\s v)$ for a unique complex structure $J_\s$ compatible with $\Omega$, and $\Psi$ is the translate by $\zeta_c$ of the vacuum of the Bargmann representation defined by $J_\s$. The map $\s\mapsto J_\s$ is a bijection from the disc $|\s|<\frac12$ onto the set of complex structures compatible with $\Omega$.
\end{proposition}

\begin{proof}
With $\s'=2\s$, \eqref{eq:aadag} gives $(\hat a-\s'\hat a^\dagger)\Psi=\beta\Psi$. The operator $\hat b=(\hat a-\s'\hat a^\dagger)/\sqrt{1-\s'\bar\s'}$ satisfies $[\hat b,\hat b^\dagger]=1$, and $\Psi$ is an eigenstate of $\hat b$, i.e.\ a coherent state for $\hat b$. Inverting, $\hat a=(\hat b+\s'\hat b^\dagger)/\sqrt{1-\s'\bar\s'}$. This gives $\langle\hat a\rangle=(\beta+\s'\bar\beta)/(1-\s'\bar\s')$ and hence \eqref{eq:centre}, since $\zeta_c=\sqrt2\langle\hat a\rangle$. For the covariances, $\langle\hat a^\dagger\hat a\rangle-|\langle\hat a\rangle|^2=\s'\bar\s'/(1-\s'\bar\s')$ and $\langle\hat a^2\rangle-\langle\hat a\rangle^2=\s'/(1-\s'\bar\s')$, so the symmetrized covariance matrix of $(\hat x,\hat p)$ is
\begin{equation*}
V=\frac1{1-\s'\bar\s'}\begin{pmatrix}\frac12|1+\s'|^2&{\rm Im}\,\s'\\{\rm Im}\,\s'&\frac12|1-\s'|^2\end{pmatrix},\qquad \det V=\frac14 .
\end{equation*}
Since $\Psi$ is a coherent state for $\hat b$, which is related to $\hat a$ by a linear canonical transformation, $\langle\hat W(\lambda,\mu)\rangle_\Psi$ is a Gaussian with this mean and covariance, and its Fourier transform is $\pi^{-1}\exp[-\frac12(r-r_c)^{\rm T}V^{-1}(r-r_c)]$ with $r=(x,p)$. Writing $\frac12V^{-1}$ in terms of $\zeta$ gives $g_\s$. A positive quadratic form of unit determinant on $\R^2$ is compatible with $\Omega$: if $G$ is its matrix and $\omega$ that of $\Omega$, then $J=-\omega G$ satisfies $J^2=-1$. Conversely, every compatible complex structure arises from exactly one $\s$, since $\s\mapsto V$ maps the disc bijectively onto the positive matrices of determinant $\frac14$.
\end{proof}

For $\s=0$ we recover the coherent states and \eqref{eq:rhocoh}; for $\beta=0$, the squeezed states $e^{\s\zeta^2}$ of \cite{ACK}, whose distributions are centred at the origin. Proposition~\ref{prop:squeezed} makes explicit the linear canonical transformation invoked in \cite{ACK}. The pure states with a classical distribution are in one-to-one correspondence with pairs $(J,\zeta_c)$: a compatible K\"ahler structure, which fixes the shape of the Gaussian, and a point of the phase space, which fixes its centre. In each case the exponent of $\rho_\Psi$ is the squared distance to $\zeta_c$ in the K\"ahler metric $g_J$, exactly as for the vacuum.

For mixed states there is no comparably simple classification. Every mixed Gaussian state has a positive distribution, with a covariance larger than that of some pure Gaussian. There are also non-Gaussian mixed states with positive distributions; for instance $(1-\epsilon)|\Psi_o\rangle\langle\Psi_o|+\epsilon|\Psi_1\rangle\langle\Psi_1|$ has $\rho=\pi^{-1}e^{-\zeta\bar\zeta}(1-2\epsilon+2\epsilon\,\zeta\bar\zeta)$, which is positive exactly when $\epsilon\le\frac12$. We refer to \cite{BrockerWerner,Mandilara} for systematic results.

\begin{figure}[t]
\centering
\includegraphics[width=\textwidth]{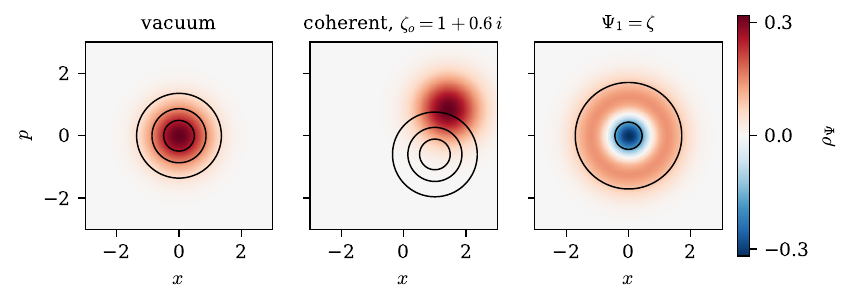}
\caption{The distribution $\rho_\Psi$ of Eq.~\eqref{eq:rhoPsi} (colour) and the Bargmann density $B_\Psi=\pi^{-1}e^{-\zeta\bar\zeta}|\Psi|^2$ (black contours at $0.05$, $0.15$, $0.25$) in the plane $\zeta=x+ip$. For the vacuum they coincide. For a coherent state they are Gaussians of the same width centred at $\sqrt2\zeta_o$ and at $\bar\zeta_o$ respectively. For $\Psi_1=\zeta$, $\rho_\Psi$ is negative near the origin, where $B_\Psi$ vanishes.}
\label{fig:densities}
\end{figure}

\subsection{What the Bargmann density represents}\label{sec:bargmanndensity}

In \cite{ACK} the distribution $\rho_o$ was introduced as the density of the Bargmann measure, i.e.\ as $|\Psi_o|^2\dmuB$ with respect to the Liouville measure. For other states the Bargmann density
\begin{equation}\label{eq:BPsi}
B_\Psi(\zeta)=\frac1\pi\,e^{-\zeta\bar\zeta}\,|\Psi(\zeta)|^2
\end{equation}
is always a probability density, but it is not $\rho_\Psi$. It does, however, reproduce a different family of expectation values exactly.

\begin{proposition}\label{prop:bargmann}
For every normalized $\Psi$, $\int_\Gamma B_\Psi\,\dmuL=1$, and
\begin{equation}\label{eq:antinormal}
\langle\Psi,\hat a^m(\hat a^\dagger)^n\Psi\rangle=\int_\Gamma\bar\zeta^m\zeta^n\,B_\Psi\,\dmuL\qquad\text{for all }m,n .
\end{equation}
Moreover
\begin{equation}\label{eq:BfromRho}
B_\Psi(\zeta)=\frac2\pi\int_\Gamma\rho_\Psi(\zeta')\,e^{-|\sqrt2\bar\zeta-\zeta'|^2}\,\dmuL(\zeta'),
\end{equation}
and $B_\Psi=\rho_\Psi$ if and only if $\Psi$ is the vacuum, up to a phase.
\end{proposition}

\begin{proof}
Since $\hat a$ is the adjoint of $\hat a^\dagger$, which acts by multiplication by $\zeta$, the left side of \eqref{eq:antinormal} is $\langle(\hat a^\dagger)^m\Psi,(\hat a^\dagger)^n\Psi\rangle=\int\bar\zeta^m\zeta^n|\Psi|^2\dmuB$. For \eqref{eq:BfromRho}, $B_\Psi(\zeta)$ is $\pi^{-1}|\langle\beta|\Psi\rangle|^2$ evaluated at the coherent label $\beta=\bar\zeta$, i.e.\ the Husimi function \cite{Husimi}. The Husimi function is the convolution of the Wigner function with the vacuum Wigner function, expressed in the variable $\beta=\zeta'/\sqrt2$ \cite{CahillGlauber}. Finally, $\int\zeta\bar\zeta\,B_\Psi\,\dmuL=\langle\hat a\hat a^\dagger\rangle=\langle\hat a^\dagger\hat a\rangle+1$, whereas $\int\zeta\bar\zeta\,\rho_\Psi\,\dmuL=\langle\hat x^2+\hat p^2\rangle=2\langle\hat a^\dagger\hat a\rangle+1$. These agree only if $\langle\hat a^\dagger\hat a\rangle=0$, i.e.\ for the vacuum, and for the vacuum $B_o=\rho_o$ by \eqref{eq:rhoo}.
\end{proof}

The two densities therefore come with two different dictionaries between operators and functions. In $\rho_\Psi$, $\hat a$ corresponds to $\zeta/\sqrt2$ and products are Weyl ordered. In $B_\Psi$, $\hat a$ corresponds to $\bar\zeta$ and products are anti-normally ordered. In the vacuum the same function serves both purposes, and this coincidence underlies the construction of \cite{ACK}. For a coherent state the two densities are Gaussians of the same width centred at different points, $\sqrt2\zeta_o$ and $\bar\zeta_o$ (Fig.~\ref{fig:densities}). For $\Psi_1$, $B_{\Psi_1}=\pi^{-1}\zeta\bar\zeta\,e^{-\zeta\bar\zeta}$ is positive, while $\rho_{\Psi_1}$ is not. The positivity of $B_\Psi$ for every state is thus bought at the price of changing the operator ordering, and the expectation values of polynomials in $\hat x$ and $\hat p$ can be recovered from $B_\Psi$ only by changing the classical functions as well. In the anti-normal dictionary $\hat x$ corresponds to $\sqrt2\,x$ and $\hat p$ to $-\sqrt2\,p$, and, for example,
\begin{equation}\label{eq:x2B}
\langle\hat x^2\rangle_\Psi=\bigl\langle 2x^2-\tfrac12\bigr\rangle_{B_\Psi},
\end{equation}
which gives $\frac12$ for the vacuum and $\frac32$ for $\Psi_1$, as it must. The general ordering calculus is given in \cite{CahillGlauber}.

\subsection{From the distribution to the Bargmann density: Gaussian smoothing}\label{sec:smoothing}

Equation \eqref{eq:BfromRho} says that the Bargmann density is a Gaussian average of $\rho_\Psi$. It is useful to interpolate. For $0\le\eta\le\frac12$ let $\rho^{(\eta)}_\Psi$ be the convolution of $\rho_\Psi$ with the centred Gaussian of covariance $\eta$ in each of $x$ and $p$. Physically, it is the distribution of the state obtained by applying random phase space displacements of that covariance, i.e.\ by adding classical noise. Alternatively, it is the distribution seen by a probe of finite resolution. The construction of Proposition~\ref{prop:rhoPsi} extends directly. With $v=1+2\eta$ and $r=-(1-2\eta)/(1+2\eta)$,
\begin{equation}\label{eq:rhoeta}
\rho^{(\eta)}_\Psi(\zeta)=\frac1{\pi v}\,e^{-\zeta\bar\zeta/v}\int_\Gamma\overline{\Psi(w)}\;e^{\sqrt2\,\zeta w/v}\;\Psi\Bigl(r\,w+\frac{\sqrt2\,\bar\zeta}v\Bigr)\dmuB(w),
\end{equation}
which reduces to \eqref{eq:rhoPsi} for $\eta=0$. It follows, exactly as in the proof of Proposition~\ref{prop:rhoPsi}, from $\rho^{(\eta)}_\Psi(\zeta)=(\pi v)^{-1}\langle\Psi,\hat D(\beta)\,r^{\hat a^\dagger\hat a}\,\hat D(\beta)^\dagger\Psi\rangle$ with $\beta=\zeta/\sqrt2$, which is the displaced-operator form of the $s$-ordered quasiprobabilities of \cite{CahillGlauber} with $s=-2\eta$, and from $r^{\hat a^\dagger\hat a}\Psi(w)=\Psi(rw)$. At $\eta=\frac12$, $r=0$ and \eqref{eq:reproducing} gives $\rho^{(1/2)}_\Psi(\zeta)=\frac12B_\Psi(\bar\zeta/\sqrt2)$: the end of the family is the Bargmann density, up to the linear change of variables in \eqref{eq:BfromRho}. Since convolution with a Gaussian preserves positivity, the set of $\eta$ for which $\rho^{(\eta)}_\Psi\ge0$ is an interval $[\eta_*,\frac12]$, and we call $\eta_*(\Psi)$ the \emph{smoothing depth} of the state. It is a Wigner-referenced version of the nonclassical depth of Lee \cite{Lee}.

\begin{proposition}\label{prop:depth}
$\eta_*=0$ exactly when $\rho_\Psi\ge0$. For a pure state there are only two possibilities: $\eta_*(\Psi)=0$ if $\Psi$ has the form \eqref{eq:gaussianPsi}, and $\eta_*(\Psi)=\frac12$ otherwise. For the mixed states $(1-\epsilon)|\Psi_o\rangle\langle\Psi_o|+\epsilon|\Psi_1\rangle\langle\Psi_1|$, $\eta_*=\max(0,\epsilon-\frac12)$.
\end{proposition}

\begin{proof}
The first statement is the definition. If a pure $\Psi$ is not of the form \eqref{eq:gaussianPsi}, then by Theorem~\ref{thm:hudson} and its proof its Bargmann function has a zero, so $B_\Psi$ vanishes somewhere. If $\rho^{(\eta)}_\Psi$ were non-negative for some $\eta<\frac12$, then $\rho^{(1/2)}_\Psi$, being a further Gaussian average of a non-negative normalized function, would be strictly positive, a contradiction. For the mixture, \eqref{eq:rhoeta} gives $\rho^{(\eta)}=(\pi v)^{-1}e^{-\zeta\bar\zeta/v}\bigl(1-2\epsilon/v+2\epsilon\,\zeta\bar\zeta/v^2\bigr)$, which is non-negative exactly when $v\ge2\epsilon$.
\end{proof}

For pure states the smoothing depth is thus a sharp dichotomy. Either the Weyl distribution is already a probability distribution, or positivity is reached only at the coarsest member of the family, the Bargmann density, which corresponds to averaging over the full width of a coherent state. In this sense the vacuum and its squeezed and displaced relatives are the only pure states for which a classical distribution with finer than coherent-state resolution exists. Mixed states can have intermediate depths, so noise or decoherence can make the transition gradual. This is relevant for interacting perturbations: a pure state that has become non-Gaussian through mode coupling has $\eta_*=\frac12$, however small its non-Gaussianity.

\subsection{Superpositions of coherent states}\label{sec:superpositions}

It was observed in \cite{ACK} that the equality of expectation values fails if a superposition of two coherent states is compared with the corresponding mixture, because the quantum expectation values are quadratic in the state and contain cross terms. With Proposition~\ref{prop:rhoPsi} the comparison can be made exact. Let
\begin{equation}\label{eq:superposition}
\Psi=\mathcal N\bigl(c_1\Psi_{\zeta_1}+c_2\Psi_{\zeta_2}\bigr),\qquad \mathcal N^{-2}=|c_1|^2+|c_2|^2+2\,{\rm Re}\bigl(\bar c_1c_2\langle\Psi_{\zeta_1},\Psi_{\zeta_2}\rangle\bigr),
\end{equation}
where $\langle\Psi_{\zeta_1},\Psi_{\zeta_2}\rangle=\exp\bigl(-\frac12\zeta_1\bar\zeta_1-\frac12\zeta_2\bar\zeta_2+\bar\zeta_1\zeta_2\bigr)$. Since coherent states are not orthogonal, the normalization involves their overlap. Then \eqref{eq:rhoPsi} and \eqref{eq:reproducing} give
\begin{equation}\label{eq:rhosuper}
\rho_\Psi=\mathcal N^2\Bigl[|c_1|^2\rho_{\Psi_{\zeta_1}}+|c_2|^2\rho_{\Psi_{\zeta_2}}+2\,{\rm Re}\bigl(\bar c_1c_2\,\rho_{12}\bigr)\Bigr],
\end{equation}
with the interference term
\begin{equation}\label{eq:rho12}
\rho_{12}(\zeta)=\frac1\pi\exp\Bigl[-\Bigl|\zeta-\frac{\zeta_1+\zeta_2}{\sqrt2}\Bigr|^2+i\Bigl(\sqrt2\,{\rm Im}\bigl((\zeta_2-\zeta_1)\bar\zeta\bigr)-{\rm Im}(\zeta_2\bar\zeta_1)\Bigr)\Bigr].
\end{equation}
The interference term is a Gaussian centred midway between the two packets, modulated by fringes whose wave vector is proportional to the separation $\zeta_2-\zeta_1$. Its integral is $\langle\Psi_{\zeta_1},\Psi_{\zeta_2}\rangle$, so it does not change the normalization appreciably when the packets are far apart, but it makes $\rho_\Psi$ negative between them.

What matters for the comparison with a mixture is which observables see the fringes. In Fourier space the answer is sharp: by \eqref{eq:weyl}, $\hat W(\lambda,\mu)$ displaces coherent states, and
\begin{equation}\label{eq:crosschar}
\bigl|\langle\Psi_{\zeta_1},\hat W(\lambda,\mu)\Psi_{\zeta_2}\rangle\bigr|=e^{-\frac12|\zeta_1-\zeta_2+\bar\alpha|^2},\qquad \alpha=\frac{\mu+i\lambda}{\sqrt2}.
\end{equation}
The cross term in $\langle\hat W(\lambda,\mu)\rangle_\Psi$ is therefore concentrated near $\bar\alpha=\zeta_2-\zeta_1$, i.e.\ at wave numbers $\sqrt{\lambda^2+\mu^2}$ of the order of the separation $D=\sqrt2|\zeta_1-\zeta_2|$ of the two packets. Let $F(x,p)=(2\pi)^{-2}\int\tilde F(\lambda,\mu)W(\lambda,\mu)\,{\rm d}\lambda\,{\rm d}\mu$ be an observable with $\Fn F=(2\pi)^{-2}\int|\tilde F|\,{\rm d}\lambda\,{\rm d}\mu<\infty$ whose Fourier transform is supported in $\lambda^2+\mu^2\le K^2$, with $K<D$. Comparing with the normalized mixture $\rho_{\rm mix}=(|c_1|^2\rho_{\Psi_{\zeta_1}}+|c_2|^2\rho_{\Psi_{\zeta_2}})/(|c_1|^2+|c_2|^2)$, \eqref{eq:rhosuper} and \eqref{eq:crosschar} give
\begin{equation}\label{eq:superbound}
\bigl|\langle F\rangle_{\rho_\Psi}-\langle F\rangle_{\rho_{\rm mix}}\bigr|\ \le\ 4\,\mathcal N^2|c_1c_2|\,\Fn F\,e^{-\frac14(D-K)^2}.
\end{equation}
Here we used $|\langle F\rangle_\rho|\le\Fn F$ for the change of normalization, and the bound \eqref{eq:crosschar} for the interference term. Observables that do not resolve phase space on the scale $1/D$ thus cannot distinguish the superposition from the mixture, up to exponentially small corrections. Polynomial observables of low degree belong to this class in practice: their Weyl expectation values are derivatives of $\langle\hat W\rangle_\Psi$ at the origin, where the cross term is of order $e^{-D^2/4}$ times a polynomial in $D$. Observables such as displaced parity, whose Weyl symbol is a $\delta$-function, see the fringes fully \cite{BanaszekWodkiewicz}. This makes precise the sense in which the failure of \cite[(5.17)]{ACK} for superpositions is, or is not, observable.
\section{Beyond Weyl ordering}\label{sec:ordering}

\subsection{An exact ordering formula}

The distribution $\rho_\Psi$ reproduces the Weyl-ordered expectation values. As emphasized in \cite{ACK}, the quantum theory contains more information, carried by products in other orderings, such as $\hat x\hat p$ or $\hat p\hat x$ as opposed to $(\hat x\hat p)_W$. For linear observables the extra information has a very simple structure. Let $\hat L_i=u_i\hat x+v_i\hat p$, $i=1,\dots,n$, with $u_i,v_i$ real or complex, and let $L_i=u_ix+v_ip$ be the corresponding functions on $\Gamma$. Their commutators are $c$-numbers, $[\hat L_i,\hat L_j]=i\{L_i,L_j\}$ with $\{L_i,L_j\}=u_iv_j-v_iu_j$. For a subset $S\subset\{1,\dots,n\}$ write $(\prod_{k\in S}\hat L_k)_W$ for the average over all orderings of the factors, and let $\mathcal P_n$ be the set of all collections $P$ of disjoint pairs $(i,j)$, $i<j$, from $\{1,\dots,n\}$, including the empty collection.

\begin{theorem}\label{thm:ordering}
For linear operators $\hat L_1,\dots,\hat L_n$,
\begin{equation}\label{eq:ordering}
\hat L_1\hat L_2\cdots\hat L_n=\sum_{P\in\mathcal P_n}\ \prod_{(i,j)\in P}\tfrac12[\hat L_i,\hat L_j]\ \Bigl(\prod_{k\notin P}\hat L_k\Bigr)_W ,
\end{equation}
where $k\notin P$ means that $k$ belongs to no pair of $P$. Consequently, for every state $\Psi$,
\begin{equation}\label{eq:orderingexp}
\langle\hat L_1\cdots\hat L_n\rangle_\Psi=\sum_{P\in\mathcal P_n}\ \prod_{(i,j)\in P}\tfrac i2\{L_i,L_j\}\ \Bigl\langle\prod_{k\notin P}L_k\Bigr\rangle_{\rho_\Psi}.
\end{equation}
\end{theorem}

\begin{proof}
Since the commutators are central, the Baker--Campbell--Hausdorff formula gives
\begin{equation*}
e^{it_1\hat L_1}e^{it_2\hat L_2}\cdots e^{it_n\hat L_n}=e^{i\sum_jt_j\hat L_j}\ \exp\Bigl(-\tfrac12\sum_{i<j}t_it_j[\hat L_i,\hat L_j]\Bigr).
\end{equation*}
The coefficient of $t_1t_2\cdots t_n$ on the left is $i^n\hat L_1\cdots\hat L_n$. On the right, the coefficient of a product $t_{k_1}\cdots t_{k_m}$ of distinct variables in $e^{i\sum t_j\hat L_j}$ is $i^m(\hat L_{k_1}\cdots\hat L_{k_m})_W$, and the second exponential contributes $\prod(-\frac12[\hat L_i,\hat L_j])$ over a collection of disjoint pairs. Comparing coefficients and using $i^{-2}=-1$ for each pair gives \eqref{eq:ordering}. Equation \eqref{eq:orderingexp} then follows from Proposition~\ref{prop:rhoPsi}.
\end{proof}

For example, $\langle\hat x\hat p\rangle_\Psi=\langle xp\rangle_{\rho_\Psi}+\frac i2$, $\langle\hat x\hat p\hat x\rangle_\Psi=\langle x^2p\rangle_{\rho_\Psi}$ because the two single pairings cancel, and
\begin{equation}\label{eq:xpxp}
\langle\hat x\hat p\hat x\hat p\rangle_\Psi=\langle x^2p^2\rangle_{\rho_\Psi}+i\,\langle xp\rangle_{\rho_\Psi}.
\end{equation}
In the last case the two double pairings cancel. Theorem~\ref{thm:ordering} makes precise the remark in \cite{ACK} that non-Weyl-ordered products differ from their classical counterparts by terms that vanish as $\hbar\to0$. In physical units each pair carries a factor $\frac{i\hbar}2\{L_i,L_j\}$. Two features are worth stressing. First, the correction terms involve only the commutators, which are state independent, and lower moments of the same distribution $\rho_\Psi$; no further information about the state is needed. Second, whether the corrections are small is not decided by $\hbar$ alone. For a pure Gaussian state the moments of $\rho_\Psi$ are themselves proportional to powers of $\hbar$ (Remark (1) of \cite[Sec.~V.A]{ACK}), so the relative size of a correction is independent of $\hbar$. What decides it is a dimensionless geometric quantity, to which we now turn.

\subsection{Three notions, one angle}\label{sec:angle}

In \cite{ACK} the fading of non-commutativity was measured by the ratio of the expectation values of the commutator and the anticommutator of the relevant observables, and squeezing by the norms of their Hamiltonian vector fields in the metric $g$. We now show that, together with the correlation coefficient of the classical distribution, these are functions of a single angle.

Let $(\Gamma,\Omega,J,g)$ be a linear K\"ahler space, with $g(u,v)=\Omega(u,Jv)$ as in \cite[Sec.~II]{ACK}, and consider the pure Gaussian state selected by $J$, possibly translated, with distribution $\rho\propto e^{-g(\gamma-\gamma_c,\gamma-\gamma_c)/\hbar}$. For a real linear function $F$ on $\Gamma$, let $X_F$ be its Hamiltonian vector field. Then, as in \cite[Sec.~IV]{ACK},
\begin{equation}\label{eq:covg}
\begin{gathered}
\sigma_F^2:=\langle(F-\langle F\rangle)^2\rangle_\rho=\frac\hbar2\,\|X_F\|^2,\qquad {\rm Cov}_\rho(F,G)=\frac\hbar2\,g(X_F,X_G),\\
\langle[\hat F,\hat G]\rangle=i\hbar\,\{F,G\}=\pm i\hbar\,\Omega(X_F,X_G).
\end{gathered}
\end{equation}
The sign in the last relation depends on conventions and plays no role below. For two linear observables define
\begin{equation}\label{eq:CdR}
C_{FG}=\frac{{\rm Cov}_\rho(F,G)}{\sigma_F\sigma_G},\qquad \delta_{FG}=\frac{|\langle[\hat F,\hat G]\rangle|}{2\sigma_F\sigma_G},\qquad R_{FG}=\frac{|\langle[\hat F,\hat G]\rangle|}{|\langle[\hat F,\hat G]_+\rangle-2\langle\hat F\rangle\langle\hat G\rangle|}.
\end{equation}
These are the correlation coefficient of $F$ and $G$ in the classical distribution, the size of the commutator relative to the classical fluctuations, and the ratio of commutator to anticommutator used in \cite[Sec.~III]{ACK}, so that $R_{FG}=\delta_{FG}/|C_{FG}|$.

\begin{lemma}\label{lem:plane}
If $u,v$ lie in a $J$-invariant real 2-plane, then $g(u,v)^2+\Omega(u,v)^2=\|u\|^2\|v\|^2$.
\end{lemma}

\begin{proof}
For $u\ne0$, the vectors $u/\|u\|$ and $Ju/\|u\|$ form a $g$-orthonormal basis of the plane, since $g(u,Ju)=\Omega(u,J^2u)=0$ and $J$ is an isometry. Writing $v=au+bJu$, one finds $g(u,v)=a\|u\|^2$ and $|\Omega(u,v)|=|g(u,Jv)|=|b|\,\|u\|^2$, while $\|v\|^2=(a^2+b^2)\|u\|^2$.
\end{proof}

\begin{proposition}\label{prop:angle}
Let $F,G$ be real linear observables whose Hamiltonian vector fields lie in a $J$-invariant plane, as is always the case for one degree of freedom, and let $\theta_{FG}\in[0,\pi]$ be the angle between $X_F$ and $X_G$ in the metric $g$. Then, in the pure Gaussian state selected by $J$,
\begin{equation}\label{eq:oneangle}
C_{FG}=\cos\theta_{FG},\qquad \delta_{FG}=\sin\theta_{FG},\qquad R_{FG}=|\tan\theta_{FG}|,\qquad \|X_F\|\,\|X_G\|\,\sin\theta_{FG}=|\{F,G\}| .
\end{equation}
In particular, for a canonical pair, $\{F,G\}=1$, the squeezing measured by $\|X_F\|\,\|X_G\|$ equals $1/\delta_{FG}$. For an arbitrary state, not necessarily Gaussian or pure, and for arbitrary linear $F,G$,
\begin{equation}\label{eq:RS}
\delta_{FG}^2\le1-C_{FG}^2 ,
\end{equation}
where $C_{FG}$ and $\sigma_F$ are computed with the moments of $\rho_\Psi$, which coincide with the symmetrized quantum moments.
\end{proposition}

\begin{proof}
The first three relations follow from \eqref{eq:covg} and Lemma~\ref{lem:plane}, since
\begin{equation*}
\cos\theta_{FG}=\frac{g(X_F,X_G)}{\|X_F\|\,\|X_G\|},\qquad \sin\theta_{FG}=\frac{|\Omega(X_F,X_G)|}{\|X_F\|\,\|X_G\|}.
\end{equation*} The last relation in \eqref{eq:oneangle} is the definition of $\sin\theta_{FG}$, written with $|\Omega(X_F,X_G)|=|\{F,G\}|$. Inequality \eqref{eq:RS} is the Robertson--Schr\"odinger uncertainty relation \cite{Robertson,Schrodinger}, $\sigma_F^2\sigma_G^2-{\rm Cov}(F,G)^2\ge\frac14|\langle[\hat F,\hat G]\rangle|^2$, divided by $\sigma_F^2\sigma_G^2$.
\end{proof}

Proposition~\ref{prop:angle} can be read as a geometric dictionary between the three notions of \cite{ACK}. The classical distribution $\rho$ correlates $F$ and $G$ to the extent that their Hamiltonian vector fields are aligned. Their commutator is small relative to their fluctuations to exactly the same extent. A canonical pair whose vector fields become nearly aligned necessarily has a large product of norms, i.e.\ is strongly squeezed. Nothing in this dictionary refers to $\hbar$. The notions remain conceptually distinct: $\rho$ exists and reproduces the Weyl-ordered $n$-point functions whatever the angle, and the angle depends on the pair of observables considered, which is why squeezing depends on the choice of canonical pair \cite[Sec.~IV, Remark 1]{ACK}. For the pair $(q_\vk,p_\vk)$ of \cite[(4.18)]{ACK}, whose vector fields are $g$-orthonormal, $\theta=\pi/2$: there is no squeezing, no correlation, and $\delta=1$, so the commutator is never negligible relative to the fluctuations. For mixed Gaussian states the classical fluctuations are larger. If the covariance is $\nu$ times that of a pure state, $\nu\ge1$, the angle relations hold with $\delta_{FG}=\sin\theta_{FG}/\nu$.

\subsection{Bounds for arbitrary orderings}

Combining Theorem~\ref{thm:ordering} with Proposition~\ref{prop:angle} gives a bound on how much an arbitrarily ordered $n$-point function can differ from the corresponding moment of $\rho$.

\begin{proposition}\label{prop:npoint}
Let $\omega$ be a quasi-free (Gaussian) state, pure or mixed, with $\langle\hat L_i\rangle_\omega=0$. Let $\hat L_1,\dots,\hat L_n$ be linear, with $\sigma_i=\sigma_{L_i}$ and $\delta_{\max}=\max_{i<j}\delta_{L_iL_j}$. Then $\langle\hat L_1\cdots\hat L_n\rangle_\omega=\langle L_1\cdots L_n\rangle_\rho=0$ for odd $n$, and for even $n$
\begin{equation}\label{eq:npointbound}
\bigl|\langle\hat L_1\cdots\hat L_n\rangle_\omega-\langle L_1\cdots L_n\rangle_\rho\bigr|\ \le\ (n-1)!!\;\frac n2\;\delta_{\max}\ \prod_{i=1}^n\sigma_i .
\end{equation}
\end{proposition}

\begin{proof}
For a quasi-free state, $\langle\hat L_1\cdots\hat L_n\rangle_\omega=\sum_\pi\prod_{(i,j)\in\pi}G_{ij}$, where the sum runs over the $(n-1)!!$ perfect pairings $\pi$ with $i<j$ in each pair, and $G_{ij}=\langle\hat L_i\hat L_j\rangle_\omega={\rm Cov}_\rho(L_i,L_j)+\frac12\langle[\hat L_i,\hat L_j]\rangle$. The classical moment is the same sum with $G_{ij}$ replaced by ${\rm Cov}_\rho(L_i,L_j)$. By \eqref{eq:RS}, $|G_{ij}|\le\sigma_i\sigma_j$ and $|{\rm Cov}_\rho(L_i,L_j)|\le\sigma_i\sigma_j$, while $|G_{ij}-{\rm Cov}_\rho(L_i,L_j)|\le\delta_{\max}\sigma_i\sigma_j$. Replacing the $n/2$ factors of each pairing one at a time bounds the difference for each pairing by $\frac n2\,\delta_{\max}\prod_i\sigma_i$.
\end{proof}

Thus the departure of any $n$-point function, in any ordering, from the classical moment is controlled by the largest angle sine among the observables involved, measured against the natural scale $\prod_i\sigma_i$ of the classical fluctuations. When all the relevant angles are small, the quantum state and the classical distribution make the same predictions for all orderings, to the stated accuracy.
\section{Fields in FLRW space-times}\label{sec:fields}

\subsection{The distribution on the covariant phase space}

We now restore $\hbar$ and return to the setting of \cite[Sec.~II]{ACK}: a test scalar field on a spatially flat FLRW space-time with metric $a^2(\eta)(-{\rm d}\eta^2+{\rm d}\vec x^{\,2})$ and spatial topology $\mathbb T^3$ of volume $V_0$. The covariant phase space $\Gamma$ of real solutions carries the symplectic structure $\Omega$. A homogeneous, isotropic Fock vacuum is specified by mode functions $e_k(\eta)$ satisfying $e_k''+2(a'/a)e_k'+k^2e_k=0$ and $e_k\bar e_k'-e_k'\bar e_k=i/a^2$. Equivalently, it is specified by the compatible complex structure $J$ and the metric $g$ \cite[Sec.~II]{ACK}. Writing a real solution as
\begin{equation}\label{eq:phiexp}
\phi(\vec x,\eta)=\frac1{\sqrt{V_0}}\sum_\vk\bigl(z_\vk\,e_k(\eta)+\bar z_{-\vk}\,\bar e_k(\eta)\bigr)e^{i\vk\cdot\vec x},\qquad \{z_\vk,\bar z_{\vk'}\}=-i\delta_{\vk,\vk'},
\end{equation}
the $z_\vk$ are Bargmann coordinates on $\Gamma$. Upon quantization $z_\vk\mapsto\hat A_\vk$ with $[\hat A_\vk,\hat A^\dagger_{\vk'}]=\hbar\delta_{\vk,\vk'}$. At any time $\eta$, the field and momentum modes are the classical functions on $\Gamma$
\begin{equation}\label{eq:modes}
\varphi_\vk(\eta)=e_k(\eta)z_\vk+\bar e_k(\eta)\bar z_{-\vk},\qquad \pi_\vk(\eta)=a^2(\eta)\bigl(e_k'(\eta)z_\vk+\bar e_k'(\eta)\bar z_{-\vk}\bigr),
\end{equation}
and $\hat\varphi_\vk(\eta)$, $\hat\pi_\vk(\eta)$ are obtained by replacing $z_\vk$ with $\hat A_\vk$ \cite[(2.29),(2.30)]{ACK}. With the real coordinates $\sqrt2z_\vk=q_\vk+ip_\vk$ of \cite[(4.18)]{ACK}, the field analogue of \eqref{eq:rhoo}, with $\hbar$ restored as in \cite[Remark (1), Sec.~V.A]{ACK}, is
\begin{equation}\label{eq:rhofield}
\rho_o=\prod_\vk\frac1{\pi\hbar}\,e^{-(q_\vk^2+p_\vk^2)/\hbar}\ ,
\end{equation}
with respect to the Liouville measure $\prod_\vk{\rm d}q_\vk{\rm d}p_\vk$. The exponent is $g(\phi,\phi)/\hbar$, since the Hamiltonian vector fields of $q_\vk$ and $p_\vk$ are $g$-orthonormal. On any finite set of modes \eqref{eq:rhofield} is an ordinary Gaussian probability density. For the whole field the density does not exist, but, as emphasized in \cite[Sec.~V.B, Remark 1]{ACK}, the measure does: the Bargmann measure $\dmuB$ itself. Equivalently, by the Bochner--Minlos theorem \cite{Minlos}, it is a Gaussian measure on a space of distributional solutions, characterized by
\begin{equation}\label{eq:charfield}
\langle e^{iF}\rangle_{\rho_o}=e^{-\frac\hbar4\|X_F\|^2}=\langle e^{i\hat F}\rangle_o
\end{equation}
for every real linear observable $F$, such as a smeared field.

The distribution \eqref{eq:rhofield} is defined on the space of solutions and does not evolve. All the time dependence is carried by the classical observables \eqref{eq:modes} through the mode functions. This is the reason why the result of \cite{ACK} is not restricted to a single instant of time.

\begin{theorem}\label{thm:alltimes}
Let $\hat O_1,\dots,\hat O_n$ be field or momentum modes $\hat\varphi_{\vk_i}(\eta_i)$, $\hat\pi_{\vk_i}(\eta_i)$, at arbitrary, not necessarily equal, times $\eta_i$, and let $O_i$ be the corresponding functions \eqref{eq:modes}. Then
\begin{equation}\label{eq:alltimes}
\langle(\hat O_1\cdots\hat O_n)_W\rangle_o=\langle O_1\cdots O_n\rangle_{\rho_o},
\end{equation}
where $(\cdot)_W$ denotes the average over all orderings of the factors. The same holds for smeared fields and momenta, and, with the appropriate Gaussian $\rho$, for every homogeneous and isotropic quasi-free state, pure or mixed. For equal times, and for quasi-free vacua and their coherent states, this is the result of \cite[Sec.~V.B and Sec.~VI]{ACK}; see also \cite[Sec.~V.A, Remark 3]{ACK} for time evolution. The extension here is to products at different times and to mixed quasi-free states.
\end{theorem}

\begin{proof}
Both sides are multilinear, so it suffices to take each $\hat O_i$ self-adjoint, using real and imaginary parts. Every such operator is of the form $\hat L=\sum_\vk(u_\vk\hat A_\vk+\bar u_\vk\hat A^\dagger_\vk)$, a finite sum. For real $t_i$, the generating function $\langle e^{i\sum_jt_j\hat O_j}\rangle_o=e^{-\frac12\langle\hat L_t^2\rangle_o}$, with $\hat L_t=\sum_jt_j\hat O_j$, generates the Weyl-ordered products by differentiation, exactly as \cite[(5.10)]{ACK}. Similarly $\langle e^{i\sum_jt_jO_j}\rangle_{\rho_o}=e^{-\frac12\langle L_t^2\rangle_{\rho_o}}$ generates the moments. Finally $\langle\hat L^2\rangle_o=\hbar\sum_\vk|u_\vk|^2=\langle L^2\rangle_{\rho_o}$, by $\langle z_\vk\bar z_\vk\rangle_{\rho_o}=\hbar/2$ and $\langle z_\vk z_{\vk'}\rangle_{\rho_o}=0$. For other homogeneous isotropic quasi-free states, the mode functions are $\tilde e_k=\alpha_ke_k+\beta_k\bar e_k$ with $|\alpha_k|^2-|\beta_k|^2=1$, and, for mixed states, $\langle z_\vk\bar z_\vk\rangle=\hbar(1+2n_\vk)/2$. The same argument applies with the corresponding Gaussian.
\end{proof}

For a pure state other than the one defined by $e_k$, the distribution is the Gaussian of the K\"ahler metric $\tilde g$ of the new complex structure. In the Bargmann coordinates of $e_k$ it is, mode by mode, a squeezed distribution of the form \eqref{eq:rhosqueezed}. For a thermal-like state with occupation numbers $n_\vk$ it is $\prod_\vk[\pi\hbar(1+2n_\vk)]^{-1}e^{-(q_\vk^2+p_\vk^2)/[\hbar(1+2n_\vk)]}$. By Theorem~\ref{thm:hudson}, applied to any finite set of modes, these Gaussian states are the only pure states for which the distribution is positive. A state with a finite number of excitations above a quasi-free state still has a distribution reproducing all Weyl-ordered $n$-point functions, but it takes negative values.

\subsection{Non-commutativity, squeezing and correlations for a mode}

Theorem~\ref{thm:ordering} applies to the operators \eqref{eq:modes}, whose commutators are $c$-numbers. Every $n$-point function of field and momentum, in any ordering and at any times, is therefore the moment of $\rho_o$ plus the explicit commutator terms \eqref{eq:orderingexp}. For example, the two-point function \cite[(2.27)]{ACK} splits as
\begin{equation}\label{eq:twopoint}
\langle\hat\phi(\vec x_1,\eta_1)\hat\phi(\vec x_2,\eta_2)\rangle_o=\langle\phi(\vec x_1,\eta_1)\phi(\vec x_2,\eta_2)\rangle_{\rho_o}+\frac{i\hbar}{V_0}\sum_\vk{\rm Im}\bigl(e_k(\eta_1)\bar e_k(\eta_2)\bigr)e^{i\vk\cdot(\vec x_1-\vec x_2)} ,
\end{equation}
the second term being half the commutator. To measure the size of these terms we apply Proposition~\ref{prop:angle} mode by mode. For $f\in\C$ let
\begin{equation}\label{eq:Ff}
F_f=f\,z_\vk+\bar f\,\bar z_{-\vk},\qquad \hat F_f=f\hat A_\vk+\bar f\hat A^\dagger_{-\vk},
\end{equation}
so that $\varphi_\vk(\eta)=F_{e_k(\eta)}$ and $\pi_\vk(\eta)=F_{a^2e_k'(\eta)}$. A direct computation with \eqref{eq:rhofield} gives
\begin{equation}\label{eq:modestats}
\sigma_f^2:=\langle F_f\bar F_f\rangle_{\rho_o}=\hbar|f|^2=\frac\hbar2\|X_{F_f}\|^2,\qquad \langle F_f\bar F_g\rangle_{\rho_o}=\hbar\,{\rm Re}(f\bar g),\qquad \langle[\hat F_f,\hat F_g^\dagger]\rangle_o=2i\hbar\,{\rm Im}(f\bar g) .
\end{equation}
The first relation reproduces \cite[(4.1)--(4.4)]{ACK}. The real and imaginary parts of $F_f$ are the observables $|f|(q_+\cos\theta_f-p_+\sin\theta_f)$ and $|f|(q_-\sin\theta_f+p_-\cos\theta_f)$ of the two independent oscillators $w_\pm=(z_\vk\pm z_{-\vk})/\sqrt2$, with $\sqrt2\,w_\pm=q_\pm+ip_\pm$ and $\theta_f=\arg f$. Their Hamiltonian vector fields are therefore rotated by $\theta_f$ within $J$-invariant planes. Proposition~\ref{prop:angle} then gives the following.

\begin{corollary}\label{cor:modes}
For two observables $F_f,F_g$ of the same mode, with $\theta_{fg}=\arg(f\bar g)$ the angle between their Hamiltonian vector fields,
\begin{equation}\label{eq:modeangle}
C_{fg}=\cos\theta_{fg},\qquad \delta_{fg}=|\sin\theta_{fg}|,\qquad R_{fg}=|\tan\theta_{fg}| ,
\end{equation}
where $C_{fg}={\rm Re}\langle F_f\bar F_g\rangle_{\rho_o}/(\sigma_f\sigma_g)$, $\delta_{fg}=|\langle[\hat F_f,\hat F_g^\dagger]\rangle_o|/(2\sigma_f\sigma_g)$ and $R_{fg}=|\langle[\hat F_f,\hat F_g^\dagger]\rangle_o|/|\langle[\hat F_f,\hat F_g^\dagger]_+\rangle_o|$. In particular, for the field and momentum at the same time, in any FLRW background,
\begin{equation}\label{eq:deltaphipi}
\delta_{\varphi\pi}(\eta)=\frac1{2a^2|e_k||e_k'|}=\frac1{\|X_{\varphi_\vk}\|\,\|X_{\pi_\vk}\|}=\frac{|R_{\varphi,\pi}|}{\sqrt{1+R_{\varphi,\pi}^2}},\qquad |R_{\varphi,\pi}(\eta)|=\frac1{|2a^2{\rm Re}(e_k\bar e_k')|},
\end{equation}
the last expression being \cite[(3.5)]{ACK}. For the field at two times,
\begin{equation}\label{eq:deltaphiphi}
\delta_{\varphi\varphi}(\eta_1,\eta_2)=\bigl|\sin\bigl(\arg e_k(\eta_1)-\arg e_k(\eta_2)\bigr)\bigr| .
\end{equation}
For smeared observables involving many modes, $\delta_{FG}\le\max_\vk\delta_{f_\vk g_\vk}$, the maximum being taken over the modes in the support.
\end{corollary}

\begin{proof}
Everything except the last statement follows from \eqref{eq:modestats} and Proposition~\ref{prop:angle}; for \eqref{eq:deltaphipi} we used the Wronskian, $a^2\,{\rm Im}(e_k\bar e_k')=\frac12$, and \cite[(4.31),(4.32)]{ACK} for the norms. For smeared observables the commutator is a sum over modes, and $|{\rm Im}(f_\vk\bar g_\vk)|\le\delta_{\max}|f_\vk||g_\vk|$; the Cauchy--Schwarz inequality then gives $\sum_\vk|f_\vk||g_\vk|\le(\sum_\vk|f_\vk|^2)^{1/2}(\sum_\vk|g_\vk|^2)^{1/2}$.
\end{proof}

\begin{figure}[t]
\centering
\includegraphics[width=\textwidth]{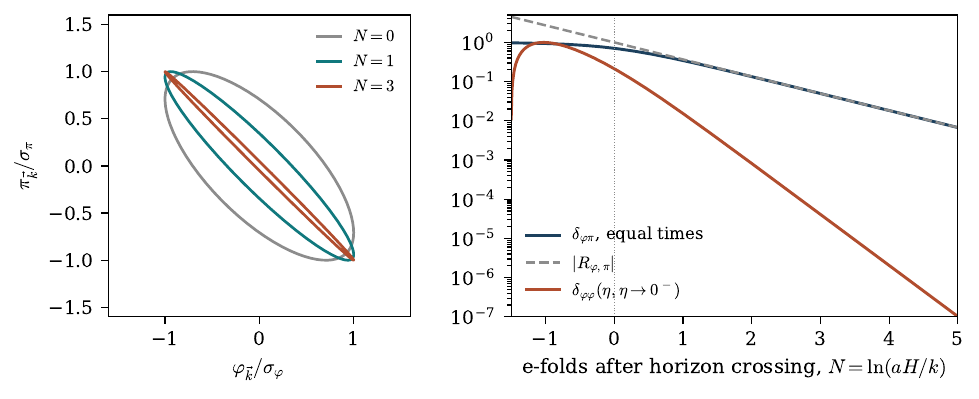}
\caption{De Sitter. Left: the covariance ellipse of $\rho_o$ in the plane of $(\varphi_\vk/\sigma_\varphi,\pi_\vk/\sigma_\pi)$ at horizon crossing and $N=1,3$ e-folds later. Its area is proportional to $\delta_{\varphi\pi}=\sin\theta$, the relative size of the commutator. Right: $\delta_{\varphi\pi}$ at equal times, the ratio $R_{\varphi,\pi}$ of \cite{ACK}, and $\delta_{\varphi\varphi}$ between a time $\eta$ and late times, as functions of the number of e-folds after horizon crossing.}
\label{fig:ellipses}
\end{figure}

Equation \eqref{eq:deltaphipi} shows explicitly that the ratio $R_{\varphi,\pi}$ used in \cite{ACK} to measure the fading of non-commutativity and the product of norms used to measure squeezing are the same quantity. Each is a function of the angle between $X_{\varphi_\vk}$ and $X_{\pi_\vk}$, and that angle is, through \eqref{eq:modeangle}, the correlation between $\varphi_\vk$ and $\pi_\vk$ in the time-independent distribution $\rho_o$. Under evolution, the vector fields of the time-dependent observables $\varphi_\vk(\eta)$ and $\pi_\vk(\eta)$ move in $\Gamma$, while $g$ and $\rho_o$ stay fixed. The fading of non-commutativity is the statement that these vector fields become aligned (or anti-aligned), so that $\rho_o$ correlates the observables perfectly and its support, as seen through them, collapses onto a line (Fig.~\ref{fig:ellipses}).

\subsection{De Sitter}

In de Sitter space-time, $a=-1/(H\eta)$, and the Bunch--Davies mode functions are \cite[(3.3)]{ACK}
\begin{equation}\label{eq:BD}
e_k=\Bigl(\frac1a+i\frac Hk\Bigr)\frac{e^{-ik\eta}}{\sqrt{2k}},\qquad a^2e_k'=-\frac{ika}{\sqrt{2k}}\,e^{-ik\eta}.
\end{equation}
Hence $\arg e_k=\Theta_k(\eta)$ with $\Theta_k=\arctan(aH/k)+k/(aH)$, and $\arg(a^2e_k')=-\frac\pi2-k\eta$. With $N=\ln(aH/k)$ the number of e-folds since horizon crossing, Corollary~\ref{cor:modes} gives
\begin{align}
\delta_{\varphi\pi}(\eta)&=\Bigl[1+\Bigl(\frac{aH}k\Bigr)^2\Bigr]^{-1/2}\simeq e^{-N},\qquad C_{\varphi\pi}=-\frac{aH}k\,\delta_{\varphi\pi}\to-1,\label{eq:dSphipi}\\
\delta_{\varphi\varphi}(\eta_1,\eta_2)&=\bigl|\sin\bigl(\Theta_k(\eta_1)-\Theta_k(\eta_2)\bigr)\bigr|\simeq\frac13\Bigl|\Bigl(\frac k{a_1H}\Bigr)^3-\Bigl(\frac k{a_2H}\Bigr)^3\Bigr|\le\frac13e^{-3N_1},\label{eq:dSphiphi}\\
\delta_{\pi\pi}(\eta_1,\eta_2)&=\bigl|\sin k(\eta_1-\eta_2)\bigr|\le e^{-N_1},\label{eq:dSpipi}
\end{align}
where $N_1$ refers to the earlier of the two times. The approximations hold after horizon crossing; we used $\Theta_k=\frac\pi2+\frac13(k/aH)^3+O((k/aH)^5)$. The first relation reproduces the result $|R_{\varphi,\pi}|=k/(aH)=e^{-N}$ of \cite[(3.6),(3.7)]{ACK}. It also shows that the product of norms $\|X_{\varphi_\vk}\|\,\|X_{\pi_\vk}\|=\sqrt{1+(aH/k)^2}$, computed from \cite[(4.31),(4.32)]{ACK}, is its inverse. The second relation gives in closed form the two-time ratio computed in \cite[App.~A]{ACK}. Since $R=\delta/|C|$, it reproduces their values: $|R|=(\sin1-\cos1)/(\sin1+\cos1)\simeq0.218$ between horizon crossing and the asymptotic future \cite[(A7)]{ACK}, and $|R|\simeq8.17\times10^{-4}$ when the earlier time is two e-folds after crossing \cite[(A8)]{ACK}. What the angle form adds is the reason and the rate. The phase of $e_k$ freezes as $(k/aH)^3$, so after horizon crossing field operators at two different times commute, relative to their fluctuations, to a much higher accuracy than field and momentum at the same time. Table~\ref{tab:rates} collects these results. By Proposition~\ref{prop:npoint} and Corollary~\ref{cor:modes}, every $n$-point function of field and momentum modes at times later than $N$ e-folds after horizon crossing, in any ordering, agrees with the corresponding moment of $\rho_o$ up to a relative error at most $(n-1)!!\,\frac n2\,e^{-N}$. If only field operators are involved, the error is at most $(n-1)!!\,\frac n6\,e^{-3N}$. For $N=55$ these numbers are of order $10^{-24}$ and $10^{-72}$ times combinatorial factors.

\begin{table}[t]
\centering
\caption{Relative size $\delta=|\sin\theta|$ of the non-commutative corrections and classical correlation $C=\cos\theta$ for pairs of mode observables in de Sitter, $N$ e-folds after horizon crossing ($N_1$ for the earlier of two times).}
\label{tab:rates}
\begin{tabular}{@{}lccc@{}}
\toprule
pair & $\delta$, exact & $\delta$, for $N\gg1$ & $C$, for $N\gg1$\\
\midrule
$\varphi_\vk(\eta)$, $\pi_\vk(\eta)$ & $[1+(aH/k)^2]^{-1/2}$ & $e^{-N}$ & $-1$\\[2pt]
$\varphi_\vk(\eta_1)$, $\varphi_\vk(\eta_2)$ & $|\sin(\Theta_k(\eta_1)-\Theta_k(\eta_2))|$ & $\le\frac13e^{-3N_1}$ & $+1$\\[2pt]
$\pi_\vk(\eta_1)$, $\pi_\vk(\eta_2)$ & $|\sin k(\eta_1-\eta_2)|$ & $\le e^{-N_1}$ & $+1$\\
$q_\vk$, $p_\vk$ of \cite[(4.18)]{ACK} & $1$ & $1$ & $0$\\
\bottomrule
\end{tabular}
\end{table}

\subsection{After inflation: the radiation era}\label{sec:radiation}

In \cite[Sec.~III]{ACK} it was found that in radiation- and matter-dominated universes the ratio $R_{\varphi,\pi}$ grows, so that non-commutativity does not fade. That analysis used the natural vacuum of each background, e.g.\ \cite[(2.22)]{ACK} for radiation. The state of perturbations in a radiation era that follows inflation is, however, the state inherited from inflation, which is highly excited relative to the radiation-era vacuum. We now show that for this state the fading persists.

\begin{figure}[t]
\centering
\includegraphics[width=\textwidth]{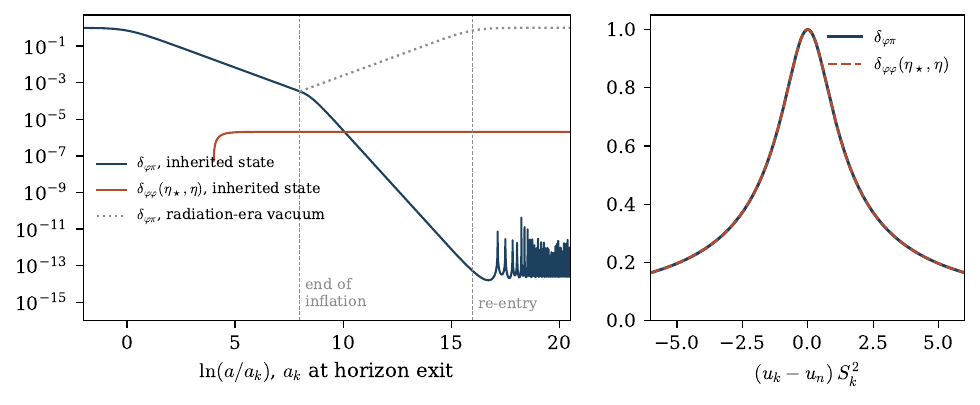}
\caption{De Sitter followed by radiation domination, for a mode leaving the Hubble radius $N_k=8$ e-folds before the end of inflation. Left: $\delta_{\varphi\pi}$ and $\delta_{\varphi\varphi}(\eta_\star,\eta)$, with $\eta_\star$ four e-folds after horizon exit, for the state inherited from inflation (solid). The radiation-era vacuum \cite[(2.22)]{ACK} is shown for comparison (dotted). After re-entry the inherited state has $\delta_{\varphi\pi}\sim1/(2n_\vk)$ with $n_\vk=e^{4N_k}/4$; the narrow spikes at the nodes are not resolved at this scale. Right: the profile of $\delta_{\varphi\pi}$ and $\delta_{\varphi\varphi}$ across a node $u_n$ of $e_k$, as a function of $u_k-u_n$ in units of $S_k^{-2}$; the two curves nearly coincide.}
\label{fig:matched}
\end{figure}

Consider de Sitter followed by radiation domination, with $a$ and $a'$ continuous at the end of inflation, $\eta=\eta_e$. Thus $a=-1/(H\eta)$ for $\eta\le\eta_e$, and $a=a_e+a_e^2H(\eta-\eta_e)$ for $\eta\ge\eta_e$. In the radiation era $\chi_k=a\,e_k$ satisfies $\chi_k''+k^2\chi_k=0$. Matching $e_k$ and $e_k'$ at $\eta_e$ to the Bunch--Davies modes \eqref{eq:BD} gives
\begin{equation}\label{eq:matching}
\chi_k=\frac{\alpha_ke^{-ik\eta}+\beta_ke^{ik\eta}}{\sqrt{2k}},\qquad \alpha_k=1-\frac i{x_k}-\frac1{2x_k^2},\qquad \beta_k=\frac{e^{-2ix_k}}{2x_k^2},\qquad x_k=k\eta_e=-e^{-N_k},
\end{equation}
where $N_k=\ln(a_eH/k)$ is the number of e-folds between horizon exit and the end of inflation, and $|\alpha_k|^2-|\beta_k|^2=1$. The occupation number with respect to the radiation-era vacuum is $n_\vk=|\beta_k|^2=e^{4N_k}/4$. Writing $\alpha_k\bar\beta_k=|\alpha_k||\beta_k|e^{2iu_0}$, one has the exact decomposition
\begin{equation}\label{eq:standing}
\chi_k=\frac{e^{i\psi_k}}{\sqrt{2k}}\Bigl[(|\alpha_k|+|\beta_k|)\cos u_k-i\,(|\alpha_k|-|\beta_k|)\sin u_k\Bigr],\qquad u_k=k\eta-u_0 ,
\end{equation}
with a constant phase $\psi_k$ and $(|\alpha_k|+|\beta_k|)(|\alpha_k|-|\beta_k|)=1$. The mode function is a standing wave whose phase is locked up to corrections of relative order $(|\alpha_k|+|\beta_k|)^{-2}\simeq1/(4n_\vk)$. Corollary~\ref{cor:modes} gives, with $S_k=|\alpha_k|+|\beta_k|$,
\begin{equation}\label{eq:radphiphi}
\delta_{\varphi\varphi}(\eta_1,\eta_2)=\frac{|\sin(u_1-u_2)|}{\sqrt{(S_k^2\cos^2u_1+S_k^{-2}\sin^2u_1)(S_k^2\cos^2u_2+S_k^{-2}\sin^2u_2)}}
\end{equation}
for two times in the radiation era. After re-entry, when $a'/a\ll k$,
\begin{equation}\label{eq:radphipi}
\delta_{\varphi\pi}(\eta)\simeq\frac1{\sqrt{(S_k^2\cos^2u_k+S_k^{-2}\sin^2u_k)(S_k^2\sin^2u_k+S_k^{-2}\cos^2u_k)}}\simeq\frac1{2n_\vk|\sin2u_k|}.
\end{equation}
Thus, away from the nodes of $\varphi_\vk$ and $\pi_\vk$, the non-commutative corrections remain suppressed by the inverse occupation number, and the phase relation established during inflation is inherited: $\delta_{\varphi\varphi}$ between a time during inflation and any later time stays at its end-of-inflation value. The exceptions are the nodes, $\cos u_k=0$ or $\sin u_k=0$. There, by \eqref{eq:radphipi}, $\delta_{\varphi\pi}=1$, and the fluctuation of the vanishing observable is so small that the commutator is no longer negligible relative to it. Each such window has width $\Delta(k\eta)\simeq S_k^{-2}\simeq1/(4n_\vk)$, so it occupies a fraction of time of order $1/n_\vk$. Figure~\ref{fig:matched} shows these features for $N_k=8$, together with the behaviour for the radiation-era vacuum of \cite{ACK}, for which $\delta_{\varphi\pi}=k\tau/\sqrt{1+k^2\tau^2}\to1$ in terms of the conformal time $\tau$ with $a\propto\tau$. For realistic $N_k\approx55$, $1/n_\vk\sim e^{-220}$.

\subsection{What the positive distribution does not capture}\label{sec:notcaptured}

A positive distribution reproducing all Weyl-ordered $n$-point functions, together with angles that make all other orderings agree to great accuracy, may suggest that nothing quantum is left. Two facts already visible for one degree of freedom show that this is not so. First, the purity of a state is $\mathrm{Tr}\,\hat\rho^2=2\pi\hbar\int\rho^2\,\dmuL$. For a pure Gaussian this equals one even though $\rho$ is a spread-out distribution: purity is not a property of $\rho$ as a classical probability distribution. Second, the Weyl symbols of quantum effects need not lie between zero and one; that of the vacuum projector is $2e^{-\zeta\bar\zeta}$ in the units of Sec.~\ref{sec:general}. In cosmology, the sharpest illustration is provided by the pair of modes $\pm\vk$.

Consider the radiation era of Sec.~\ref{sec:radiation}, and let $\hat c_{\pm\vk}$ be the annihilation operators of the radiation-era vacuum \cite[(2.22)]{ACK}, defined by $\hat\varphi_\vk=f_k\hat c_\vk+\bar f_k\hat c^\dagger_{-\vk}$ with $f_k=e^{-ik\eta}/(a\sqrt{2k})$. Comparing with $e_k=\alpha_kf_k+\beta_k\bar f_k$ from \eqref{eq:matching} gives $\hat c_\vk=\alpha_k\hat A_\vk+\bar\beta_k\hat A^\dagger_{-\vk}$. In the Bargmann representation built on $\hat c_{\pm\vk}$, with $\hat c^\dagger_{\pm\vk}$ acting as multiplication by $\zeta_{\pm\vk}$, the condition $\hat A_\vk\Psi=0$ reads $(\bar\alpha_k\partial_{\zeta_\vk}-\bar\beta_k\zeta_{-\vk})\Psi=0$, and similarly with $\vk\leftrightarrow-\vk$. Hence
\begin{equation}\label{eq:tmsv}
\Psi(\zeta_\vk,\zeta_{-\vk})=\Bigl(1-\frac{|\beta_k|^2}{|\alpha_k|^2}\Bigr)^{1/2}\exp\Bigl(\frac{\bar\beta_k}{\bar\alpha_k}\,\zeta_\vk\zeta_{-\vk}\Bigr),
\end{equation}
which is of the Gaussian form of Theorem~\ref{thm:hudson} for two degrees of freedom, with an off-diagonal quadratic term. Its distribution is positive, as it must be. But \eqref{eq:tmsv} is a two-mode squeezed vacuum, with squeezing parameter $r_k$ given by $\sinh^2r_k=|\beta_k|^2=n_\vk$. In the factorization defined by $\hat c_\vk$ and $\hat c_{-\vk}$, the modes $\vk$ and $-\vk$ are therefore entangled. The reduced state of one of them is thermal with mean occupation $n_\vk$, and the entanglement entropy is
\begin{equation}\label{eq:entropy}
S_\vk=(n_\vk+1)\ln(n_\vk+1)-n_\vk\ln n_\vk\simeq\ln n_\vk+1\simeq4N_k-\ln4+1 .
\end{equation}
The corresponding statements for the positive distribution are different. The classical mutual information of $\rho_o$ between the two modes, computed in the same coordinates, is $I_\rho=2\ln\cosh2r_k=2\ln(1+2n_\vk)\simeq8N_k-2\ln2$. The quantity to compare it with is the quantum mutual information of the pair, $2S_\vk\simeq8N_k+2-4\ln2$, not the entanglement entropy $S_\vk$, which is half of it. Even so, $I_\rho$ is not $2S_\vk$ but its R\'enyi-2 counterpart. For a Gaussian state with covariance $V$, $-\ln\mathrm{Tr}\,\hat\rho^2=\frac12\ln\det(2V/\hbar)$, so $I_\rho=S_2(\hat\rho_\vk)+S_2(\hat\rho_{-\vk})-S_2(\hat\rho)$ with $S_2=-\ln\mathrm{Tr}\,\hat\rho^2$ \cite{AGS}. Since R\'enyi entropies do not increase with their order, $I_\rho<2S_\vk$ for every $N_k$; both grow as $8N_k$, and their difference tends to $2-2\ln2\simeq0.614$. The entanglement is detected by linear observables through the separability criteria of \cite{Duan,Simon}. For the quadratures $\hat c_{\pm\vk}=(\hat q^c_{\pm\vk}+i\hat p^c_{\pm\vk})/\sqrt{2}$, after a local phase rotation fixed by $\arg(\alpha_k\beta_k)$,
\begin{equation}\label{eq:duan}
{\rm Var}(\hat q^c_\vk-\hat q^c_{-\vk})+{\rm Var}(\hat p^c_\vk+\hat p^c_{-\vk})=2\hbar\,e^{-2r_k}<2\hbar ,
\end{equation}
whereas every separable state gives at least $2\hbar$. Equivalently, partial transposition maps the covariance of $\rho_o$, which is positive as a classical covariance, into one with symplectic eigenvalue $\frac\hbar2e^{-2r_k}$, violating the quantum bound; the logarithmic negativity is $2r_k\simeq4N_k$, since $r_k={\rm arcsinh}\sqrt{n_\vk}\simeq2N_k$.

Bell inequalities provide a sharper test. Quadrature measurements on the two modes admit a local hidden-variable model built from the positive distribution itself: sample a point of $\Gamma$ from $\rho_o$ and evaluate the local linear functions. Displaced-parity measurements $(-1)^{\hat c^\dagger\hat c/\hbar}$, whose correlations are values of the Wigner function \cite{BanaszekWodkiewicz,BW99}, do not admit one. The CHSH combination for \eqref{eq:tmsv} depends only on $r_k$. Restricting the displacements to one quadrature, with settings $0,d$ for one mode and $0,-d$ for the other (after the local phase rotation above), gives the closed form $1+2e^{-cx}-e^{-2(c+s)x}$ with $x=d^2/\hbar$, $c=\cosh2r_k$ and $s=\sinh2r_k$. It is maximized at $x=\ln[(c+s)/c]/(c+2s)$. This exceeds the local bound $2$ for every $r_k>0$ and tends to $1+3\cdot2^{-4/3}\simeq2.191$ \cite{BanaszekWodkiewicz}. Optimizing numerically over all displacements gives larger values, Table~\ref{tab:bell}, approaching $8\cdot3^{-9/8}\simeq2.324$ \cite{Jeong}. Since $r_k\simeq2N_k$, a mode a few e-folds outside the Hubble radius at the end of inflation already reaches this asymptotic violation. In cosmology, Bell inequalities of this and related types have been analysed by Martin and Vennin \cite{MartinVennin}.

\begin{table}[t]
\centering
\caption{Maximal CHSH value for displaced-parity measurements on the two-mode squeezed vacuum \eqref{eq:tmsv}, as a function of $r_k$: restricted settings (closed form) and numerical optimization over all four displacements.}
\label{tab:bell}
\begin{tabular}{@{}ccccccc@{}}
\toprule
$r_k$ & $0.25$ & $0.5$ & $1$ & $2$ & $3$ & $\infty$\\
\midrule
restricted & $2.080$ & $2.144$ & $2.184$ & $2.190$ & $2.191$ & $1+3\cdot2^{-4/3}$\\
all displacements & $2.107$ & $2.220$ & $2.308$ & $2.324$ & $2.324$ & $8\cdot3^{-9/8}$\\
\bottomrule
\end{tabular}
\end{table}

These results are fully compatible with those of the previous subsections. The occupation number $n_\vk$ that makes the non-commutative corrections negligible for linear observables, $\delta_{\varphi\pi}\sim1/(2n_\vk)$ by \eqref{eq:radphipi}, is the same number that makes the entanglement entropy large and the Bell violation maximal. Large squeezing makes the linear observables classical and, at the same time, makes non-linear observables with singular Weyl symbols maximally non-classical. Entanglement between $\vk$ and $-\vk$ is also a statement about a factorization. In the factorization defined by $\hat A_{\pm\vk}$ the state is the vacuum, a product state, so the entanglement measures the relative position of the complex structure $J$ selecting the state and the complex structure of the radiation-era vacuum used to define the subsystems. It is, moreover, entanglement between global Fourier modes, which local observers cannot access \cite{RAB}. For finitely many field modes of compact support, whose reduced states are always mixed, entanglement is much less common than for regions of space, already in the Minkowski vacuum \cite{Ubiquitous}. Ribes-Metidieri, Agullo and Bonga have shown, using the metric and complex structure that the vacuum induces on the classical phase space, that in the Bunch--Davies vacuum of de Sitter space such modes are more correlated but less entangled than in the Minkowski vacuum, and that their entanglement decreases as the Hubble rate grows \cite{RAB,RABlong}. The large entanglement of the pair $\pm\vk$ therefore does not imply entanglement between local observables. None of this affects the equality of Weyl-ordered $n$-point functions established in \cite{ACK}; it delimits what that equality means.

\subsection{Remarks}

(1) \emph{What is time independent.} Theorem~\ref{thm:alltimes} holds in any FLRW background, with or without inflation, because $\rho_o$ lives on the space of solutions. What depends on the dynamics is the geometry of the observables $\varphi_\vk(\eta),\pi_\vk(\eta)$ relative to $\rho_o$, i.e.\ the angles of Corollary~\ref{cor:modes}. Inflation drives these angles to $0$ or $\pi$, and the subsequent radiation era preserves this alignment for the inherited state, except near nodes. This is the precise sense in which inflation, though not necessary for the existence of the distribution, is what makes the remaining non-Weyl-ordered information negligible.

(2) \emph{Relation to the decaying mode.} The alignment of the vector fields means that, after horizon crossing, $\varphi_\vk(\eta)$ and $\pi_\vk(\eta)$ measure essentially the same linear combination of $z_\vk$ and $\bar z_{-\vk}$: the amplitude of the growing mode. The orthogonal combination, the decaying mode, contributes to both observables with a weight suppressed by powers of $k/(aH)$. This is the geometric content of the familiar statement that, after horizon crossing, the perturbations behave as a classical stochastic field whose realizations differ only in the amplitude of the growing mode \cite{PolarskiStarobinsky,KPS}. Here it appears as a statement about $\rho_o$ and the angles of Corollary~\ref{cor:modes}, with explicit error bounds, and without an appeal to decoherence.

(3) \emph{Mixed states and interactions.} For mixed quasi-free states, for instance a thermal pre-inflationary state, all angle relations hold with $\delta$ divided by $1+2n_\vk$, so classical noise further suppresses the relative size of the commutators. Interactions, i.e.\ mode coupling, change the picture in two ways. The state generally ceases to be Gaussian, so that, by Theorem~\ref{thm:hudson}, its distribution need not stay positive. And the Weyl symbols of evolved observables acquire quantum corrections. These effects are beyond the scope of this paper.
\section{Discussion}\label{sec:discussion}

The analysis of \cite{ACK} showed that the quantum vacuum of linear perturbations can be replaced, for the purpose of computing Weyl-ordered $n$-point functions, by a classical distribution $\rho_o$ on phase space, and that this replacement is exact. In this paper we have asked how far the replacement extends and what, precisely, it leaves out.

On the first question, the construction extends to every state: Eq.~\eqref{eq:rhoPsi} gives, directly in the Bargmann representation, a real normalized function $\rho_\Psi$ whose moments are the Weyl-ordered expectation values. It is a probability distribution, for pure states, exactly for the coherent and squeezed states considered in \cite{ACK}. Each of these is the Gaussian of a compatible K\"ahler structure centred at a point of phase space, so that the geometric picture of \cite{ACK} is not tied to a particular vacuum. The Bargmann density $|\Psi|^2\dmuB$, which provided $\rho_o$ in \cite{ACK}, is always positive but reproduces anti-normally ordered products, and coincides with $\rho_\Psi$ only for the vacuum. For superpositions of coherent states the comparison with the corresponding mixture is decided by the resolution of the observable. Observables coarser than the separation of the packets cannot tell them apart; displaced parity and other observables with singular Weyl symbols can.

On the second question, for linear observables the answer is complete and simple. Every ordering differs from the Weyl ordering by sums of products of half-commutators, which are $c$-numbers, multiplied by lower moments of the same distribution, Eq.~\eqref{eq:orderingexp}. The only quantum information not contained in $\rho_\Psi$ is therefore the symplectic structure itself. Its importance relative to the classical fluctuations is measured, for each pair of observables, by an angle in the K\"ahler geometry of the state. The cosine of that angle is the correlation coefficient in $\rho$, its sine the relative size of the commutator, its tangent the ratio $R$ of \cite{ACK}, and for a canonical pair its cosecant the squeezing, Proposition~\ref{prop:angle}. In this sense the three notions of classicality discussed in \cite{ACK} are three readings of one geometric quantity. They remain conceptually distinct, as emphasized there: $\rho$ exists whatever the angle, and the angle depends on which observables are considered. But they are no longer independent. In particular, the smallness of non-Weyl-ordered corrections is not a consequence of $\hbar\to0$; for pure Gaussian states it is independent of $\hbar$. It is a consequence of the dynamics, which aligns the Hamiltonian vector fields of the observables of interest with respect to a fixed K\"ahler metric.

For fields in FLRW space-times, the distribution lives on the space of solutions and does not evolve, so the equality of Weyl-ordered $n$-point functions holds at arbitrary, possibly different, times, and for every homogeneous isotropic quasi-free state. The angles evolve. In de Sitter the commutator of field and momentum at equal times becomes negligible as $e^{-N}$ relative to their fluctuations, and the commutator of the field at two times as $e^{-3N}/3$. Proposition~\ref{prop:npoint} turns these rates into bounds on arbitrarily ordered $n$-point functions. After inflation, the state inherited by a radiation era keeps the alignment, with relative corrections of the order of the inverse occupation number, except within short windows around the nodes of the mode functions. This complements the finding of \cite{ACK} that non-commutativity does not fade in the natural vacuum of a radiation-dominated universe: what matters is the state, not the background alone.

We have not addressed the interpretation of the classical ensemble, i.e.\ whether and how one of its members is realized, which has been the subject of debate \cite{BOS}. The results here concern which predictions of the quantum state coincide with those of the classical distribution, for which observables, and to what accuracy. Section~\ref{sec:notcaptured} makes the limits explicit. The same squeezing that makes linear observables classical makes the pair $\pm\vk$ strongly entangled in the factorization defined by the radiation-era vacuum, with entropy $\simeq4N_k$. This is entanglement between global modes; between field modes of compact support the Bunch--Davies state carries less entanglement than the Minkowski vacuum \cite{RAB,RABlong}. It also produces Bell violations for displaced-parity measurements, which also detect the interference terms of Sec.~\ref{sec:superpositions}, close to the maximal value $8\cdot3^{-9/8}$ attainable with them \cite{Jeong,MartinVennin}. These observables are not linear, and their Weyl symbols are singular, so they lie outside the scope of Proposition~\ref{prop:angle}. The coincidence of predictions established here is a statement about a specified family of observables, not about all measurements. Similarly, the smoothing depth of Sec.~\ref{sec:smoothing} shows that, for pure states, a positive distribution finer than the coherent-state resolution exists only in the Gaussian case. Any non-Gaussianity of a pure state, however small, pushes positivity all the way to the Bargmann density. It would be interesting to extend the analysis to interacting fields, where the state ceases to be Gaussian and the Weyl symbols of evolved observables acquire quantum corrections. The exact ordering formula and the angle bounds of this paper should provide a useful starting point for such an analysis.

\section*{Acknowledgments}
I would like to thank Abhay Ashtekar and Aruna Kesavan for the collaboration on which this work builds.

\appendix
\section{Numerical checks}\label{app:checks}

All closed-form results were checked numerically in a truncated Fock basis of dimension 60--70. Formula \eqref{eq:rhoPsi} was compared, at several points, with the displaced-parity expression for a random state supported on the first five levels, and the Weyl-ordered moments up to order five with the moments of $\rho_\Psi$ computed by quadrature; the agreement is at the level of $10^{-9}$ or better. The ordering formula \eqref{eq:ordering} was checked for random products of up to five linear operators in the same non-Gaussian state, and the squeezed-state formulas \eqref{eq:rhosqueezed}--\eqref{eq:centre}, the anti-normal rule \eqref{eq:antinormal}, the relation \eqref{eq:BfromRho}, the superposition formulas \eqref{eq:rhosuper}--\eqref{eq:crosschar} and the equality case of \eqref{eq:RS} were all verified to machine precision. Inequality \eqref{eq:npointbound} was tested on 200 random sets of up to six linear observables in a squeezed state; the bound is saturated for $n=2$ and never exceeded. The smoothed formula \eqref{eq:rhoeta} was compared with the generalized-parity expression and with direct Gaussian convolution, the four-point test of Sec.~\ref{sec:rhoPsi} with the Weyl operators computed in the Fock basis, and the two-mode state \eqref{eq:tmsv} with the condition $\hat A_\vk\Psi=0$, its reduced occupation, its entropy and the R\'enyi-2 identity for $I_\rho$. The CHSH values of Table~\ref{tab:bell} were obtained by optimizing over all four displacements with several hundred random initial conditions. The de Sitter formulas \eqref{eq:dSphipi}--\eqref{eq:dSpipi} were compared with direct numerical integration of the mode equation from deep inside the Hubble radius. The matched model of Sec.~\ref{sec:radiation} was evaluated with the exact decomposition \eqref{eq:standing}, which avoids the cancellations between the large coefficients $\alpha_k$ and $\beta_k$, and validated against direct evaluation for $N_k=3$. Scripts that check the closed forms and the numbers quoted in the text, and that generate the figures, are included as ancillary files.

\end{document}